\documentclass[twocolumn,amsmath,amssymb,nofootinbib,superscriptaddress,longbibliography]{revtex4-2}

\usepackage{graphicx}
\usepackage{amsmath}
\usepackage{amssymb}
\usepackage{xcolor}
\usepackage{hyperref}
\hypersetup{pdfstartview={FitH},pdfpagemode={UseNone}, breaklinks=true, colorlinks=true,bookmarks=false,linkcolor=blue, citecolor=blue, urlcolor=blue, bookmarksopen=true, pdfnewwindow=true}
\usepackage[all]{hypcap}
\usepackage[english]{babel}
\usepackage{physics}
\usepackage{braket}

\begin{document}

\preprint{AIP/123-QED}

\title{Topological edge states in all-dielectric square-lattice arrays of bianisotropic microwave resonators}

\author{Alina~D.~Rozenblit}
  \email{alina.rozenblit@metalab.ifmo.ru}
\affiliation{School of Physics and Engineering, ITMO University, 49 Kronverksky pr., bldg. A, 197101 Saint Petersburg, Russia}

\author{Georgiy~D.~Kurganov}
\affiliation{School of Physics and Engineering, ITMO University, 49 Kronverksky pr., bldg. A, 197101 Saint Petersburg, Russia}

\author{Dmitry~V.~Zhirihin}
\affiliation{School of Physics and Engineering, ITMO University, 49 Kronverksky pr., bldg. A, 197101 Saint Petersburg, Russia}

\author{Nikita~A.~Olekhno}
\affiliation{School of Physics and Engineering, ITMO University, 49 Kronverksky pr., bldg. A, 197101 Saint Petersburg, Russia}

\date{\today}

\begin{abstract}
We demonstrate that a bianisotropic response associated with a broken mirror symmetry of a dielectric resonator allows opening a band gap in simple square lattice arrays of such resonators. Realizing the proposed system as an array of high-index ceramic resonators working at GHz frequencies, we numerically and experimentally demonstrate the presence of topological edge states at the interface between two domains with opposite orientations of the bianisotropic resonators, as well as at the boundary between a single domain and free space. For both cases, we experimentally characterize the dispersion of edge states, and we examine their propagation along sharp bends, their resilience to various types of geometrical defects, and a spin-momentum-locked unidirectional propagation in the case of circularly polarized excitation. Also, we develop a theoretical model based on a Green's function approach that describes the square lattice of resonators and features quadratic degeneracies in the vicinity of $\Gamma$ and $M$ high-symmetry points that are removed upon the introduction of bianisotropy, and apply this model to evaluate Berry curvature. The considered design opens possibilities in the construction of optical and microwave structures simultaneously featuring topological edge states at the interfaces between distinct resonator domains or a resonator domain and free space.
\end{abstract}

\maketitle

\section{Introduction}
\label{sec:Introduction}

Edge and surface states propagating along the interfaces of media with distinct topological properties~\cite{2014_Lu, 2019_Ozawa} demonstrate a set of fascinating physical phenomena related to their topological protection, including bulk-boundary correspondence~\cite{2016_Silveirinha}, resilience to geometrical imperfections in the structure~\cite{2019_Kruk}, and the absence of backscattering at sharp bends~\cite{2009_Wang, 2016_Slobozhanyuk, 2019_Slobozhanyuk, 2023_Wang}.

Such states have been experimentally demonstrated on various platforms, including microwave systems~\cite{2018_Li}, electrical circuits with passive~\cite{2019_Serra_Garcia} or active components~\cite{2021_Kotwal}, and mechanical systems~\cite{2015_Susstrunk}, including those with actively moving particles~\cite{2022_Shankar} or robots~\cite{2020_Yang}. Among recent trends are reconfigurable platforms in acoustics~\cite{2020_Darabi} and infrared optics~\cite{2024_On}. Moreover, photonic topological edge states have promising practical applications at frequencies from microwave to optical. Particular examples include leaky wave antennas based on complementary rhombic~\cite{2024_Abtahi} and hexagonal~\cite{2020_Lumer} lattices, miniature on-chip phase shifters~\cite{2022_Wang}, the demonstration of photonic chip-based beamformers for 6G wireless networks~\cite{2024_Wang}, and the enhancement of the radiation efficiency of a waveguide to free space in the THz domain~\cite{2023_Jia}.

\begin{figure}[b]
    \includegraphics[width=8.5cm]{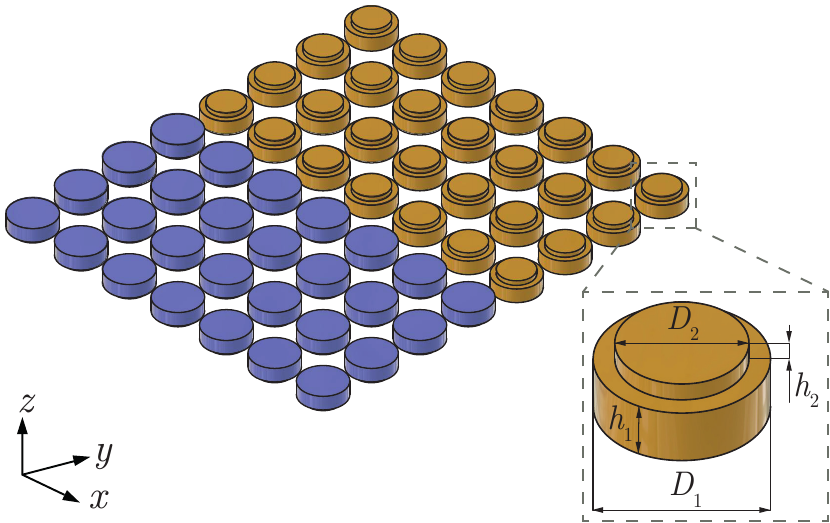}
    \caption{The schematics of the proposed photonic topological insulator including an interface between two arrays of oppositely oriented bianisotropic resonators forming a square lattice. The color of the bianisotropic resonators highlights their orientation. The inset shows an individual resonator with characteristic dimensions specified.}
    \label{fig:System}
\end{figure}

Recently, setups based on coupled all-dielectric resonators became widespread in microwave~\cite{2019_Gorlach, 2020_Li, 2022_Kurganov} and optical~\cite{2019_Kruk} experiments owing to their low loss levels, with high-index ceramic resonators being a prominent example in microwaves. One of the approaches to implement topological phenomena in arrays of such resonators utilizes their individual bianisotropic response, i.e., a coupling of the electric and magnetic moments, which can be introduced by breaking the inversion symmetry of the resonator's shape~\cite{2015_Alaee, 2015_Mousavi, 2018_Asadchy} and leads to the formation of hybrid resonances. In turn, this hybridization of the electric and magnetic modes represents the analog of the spin-orbit interaction in quantum systems and leads to the opening of a band gap~\cite{2016_Slobozhanyuk, 2019_Gorlach, 2021_Bobylev}.

\begin{figure*}[t]
    \includegraphics[width=16cm]{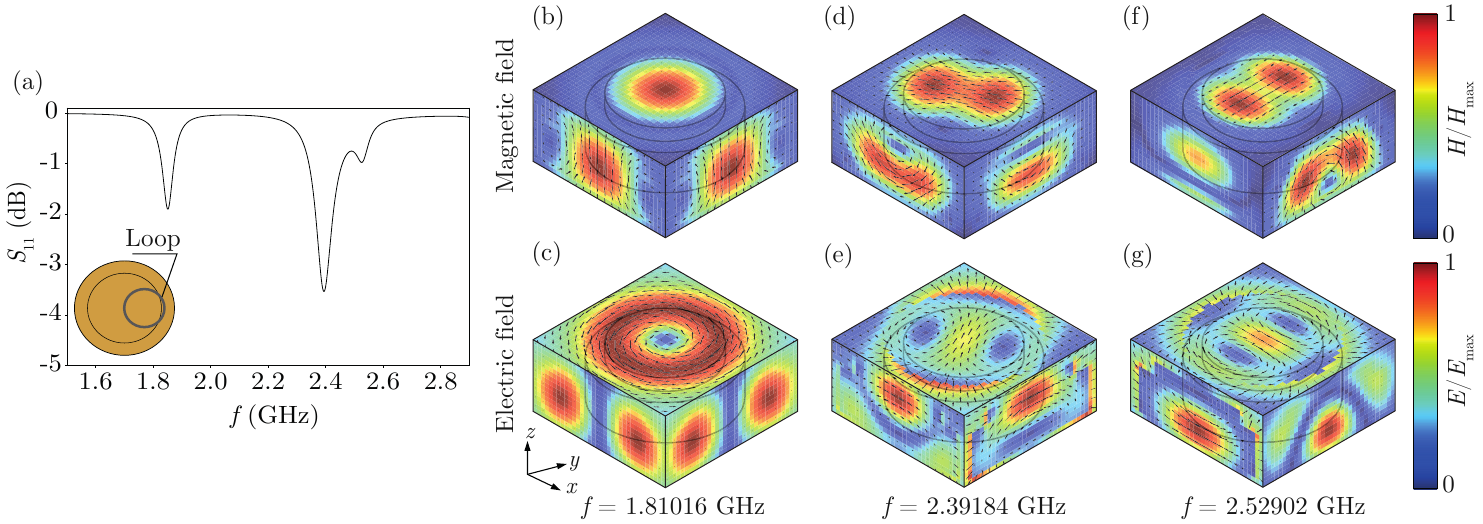}
    \caption{(a) Numerically calculated $S_{11}$ parameter for an individual bianisotropic resonator. The inset shows the position of the loop antenna in the numerical model. (b)-(g) Magnetic and electric field distributions at the three orthogonal central cross-sections of the bianisotropic resonator at the frequencies of (b),(c) $f = 1.81016$~GHz, (d),(e) $f = 2.39184$~GHz, and (f),(g) $f = 2.52902$~GHz. For visual clarity, the cross sections are translated to the facets of a box enclosing the resonator. The color shows the normalized absolute value of the magnetic or electric fields, while the black arrows visualize the tangential components of the magnetic or electric fields in the respective planes.}
    \label{fig:CST_Resonator_fields}
\end{figure*}

Although two-dimensional~\cite{2019_Slobozhanyuk} (2D) and three-dimensional~\cite{2017_Slobozhanyuk} (3D) hexagonal lattices of dielectric bianisotropic particles featuring Dirac dispersion are well studied, the systems with such particles arranged in a square lattice remain experimentally unexplored, with only lossy metallic counterparts at hand~\cite{2016_Slobozhanyuk}. In addition, edge- and corner-localized topological states have been demonstrated in square lattices with unconventionally selected unit cell~\cite{2022_Chen}, including the case where the lattice is formed by magneto-optic rods~\cite{2024_Lan}. Recently, the 2D Su-Schrieffer-Heeger model realized as an equidistant square lattice of bianisotropic resonators with staggered orientations was studied numerically and theoretically~\cite{2024_Sang}. Moreover, square-lattice systems corresponding to different topological models are extensively realized as electrical circuits~\cite{2020_Olekhno, 2022_D4, 2019_Liu} and photonic setups~\cite{2018_Chen, 2019_Ota, 2019_Xie, 2019_Chen, 2020_Han}.

In this article, we study experimentally and theoretically a 2D array of ceramic bianisotropic resonators placed at the sites of a simple square lattice that supports edge-localized states; see Fig.~\ref{fig:System}. The introduced system is essentially a single layer of the model theoretically considered in Ref.~\cite{2017_Ochiai}. We demonstrate the presence of edge states at the interface between two lattice domains with oppositely oriented resonators, as well as at the interface between a single lattice domain and free space. Moreover, we study spin-momentum locking of such states in the case of a circularly polarized excitation and show that they propagate robustly along the sharp bends.

The paper is organized as follows. In Sec.~\ref{sec:Structure}, we study numerically an individual bianisotropic resonator and discuss numerical simulations of an infinite 2D square lattice of resonators. Then, in Sec.~\ref{sec:Experiments}, we present experimental results for edge states at the interface between two bianisotropic domains as well as at the free space-single domain interface, considering their dispersion, pseudospin-momentum locking, and propagation along sharp bends. The effective Hamiltonian and topological properties of the model considered are analyzed in Sec.~\ref{sec:Hamiltonian}. Section~\ref{sec:Discussion} contains a discussion of the results and an outlook.

\section{The structure and properties of a bianisotropic planar array}
\label{sec:Structure}

We start by studying the properties of a single bianisotropic resonator with the CST Microwave Studio 2022 software package. The resonator is implemented as two concentrically attached cylinders with diameters $D_{1} = 29.1$~mm and $D_{2} = 22$~mm and heights $h_{1} = 9$~mm and $h_{2} = 3$~mm, respectively; see Fig.~\ref{fig:System}. The symmetry axes of the cylinders are aligned with the $z$ axis of the coordinate system. The permittivity of the resonator material is set to $\varepsilon = 39$, while the loss angle is $\delta = 0.0001$ at the frequency $f = 2.43$~GHz. The resonator is located in vacuum with $\varepsilon=1$.

To characterize numerically the resonant spectrum of the bianisotropic resonator, we introduce a loop antenna with diameter $12$~mm implemented as a torus made of a $0.4$~mm thick perfect electric conductor wire and placed atop the resonator at height of $1$~mm, as shown in the inset of Fig.~\ref{fig:CST_Resonator_fields}(a). Then, we evaluate the frequency dependence of the $S_{11}$ parameter using the frequency domain solver. As seen from the spectrum obtained in Fig.~\ref{fig:CST_Resonator_fields}(a), the resonator features resonances at frequencies close to $f = 1.85$~GHz, $f = 2.39$~GHz, and $f = 2.53$~GHz. The same procedure is experimentally implemented for several resonators with the same dimensions and material properties to characterize their spectra, as discussed in the Supplemental Material~\cite{Supplement}.

Next, with the help of the eigenvalue solver, we visualize the spatial distributions of the magnetic and electric field amplitudes in three orthogonal planes passing through the center of the resonator for each of the discussed resonances; see Figs.~\ref{fig:CST_Resonator_fields}(b)-\ref{fig:CST_Resonator_fields}(g). For visual clarity, the corresponding maps are shifted to the facets of the box that incorporate the resonator. Each of the field profiles is normalized to the maximum value of its amplitude among the three planes. The directions and in-plane amplitudes of tangential parts of the magnetic and electric fields in the corresponding planes are shown with black arrows, while the color denotes the total magnitude of the respective field.

The eigenstate at the frequency $f = 1.81016$~GHz clearly represents a magnetic dipole, i.e., a current loop in the $(xy)$ plane; see Figs.~\ref{fig:CST_Resonator_fields}(b)~and~\ref{fig:CST_Resonator_fields}(c). In particular, it is seen that the $z$ component of the magnetic field is maximal and nearly uniform at the symmetry axis of the resonator while the $z$ component of the electric field vanishes. In turn, the in-plane $(xy)$ component of the electric field is directed tangentially to the cylinder surface, characteristic of a current loop in the $(xy)$ plane. At higher frequencies $f = 2.39184$~GHz and $f = 2.52902$~GHz, the eigenstates shown in Figs.~\ref{fig:CST_Resonator_fields}(d),~\ref{fig:CST_Resonator_fields}(e) and Figs.~\ref{fig:CST_Resonator_fields}(f),~\ref{fig:CST_Resonator_fields}(g) render as combinations of electric and magnetic dipoles directed in the $(xy)$ plane. This becomes evident when studying the projections of the $x$, $y$, and $z$ components of the electric and magnetic fields separately; see a detailed analysis of the eigenmodes for the individual resonator in Supplemental Material Note~1~\cite{Supplement}.

\begin{figure}[tbp]
    \includegraphics[width=8.5cm]{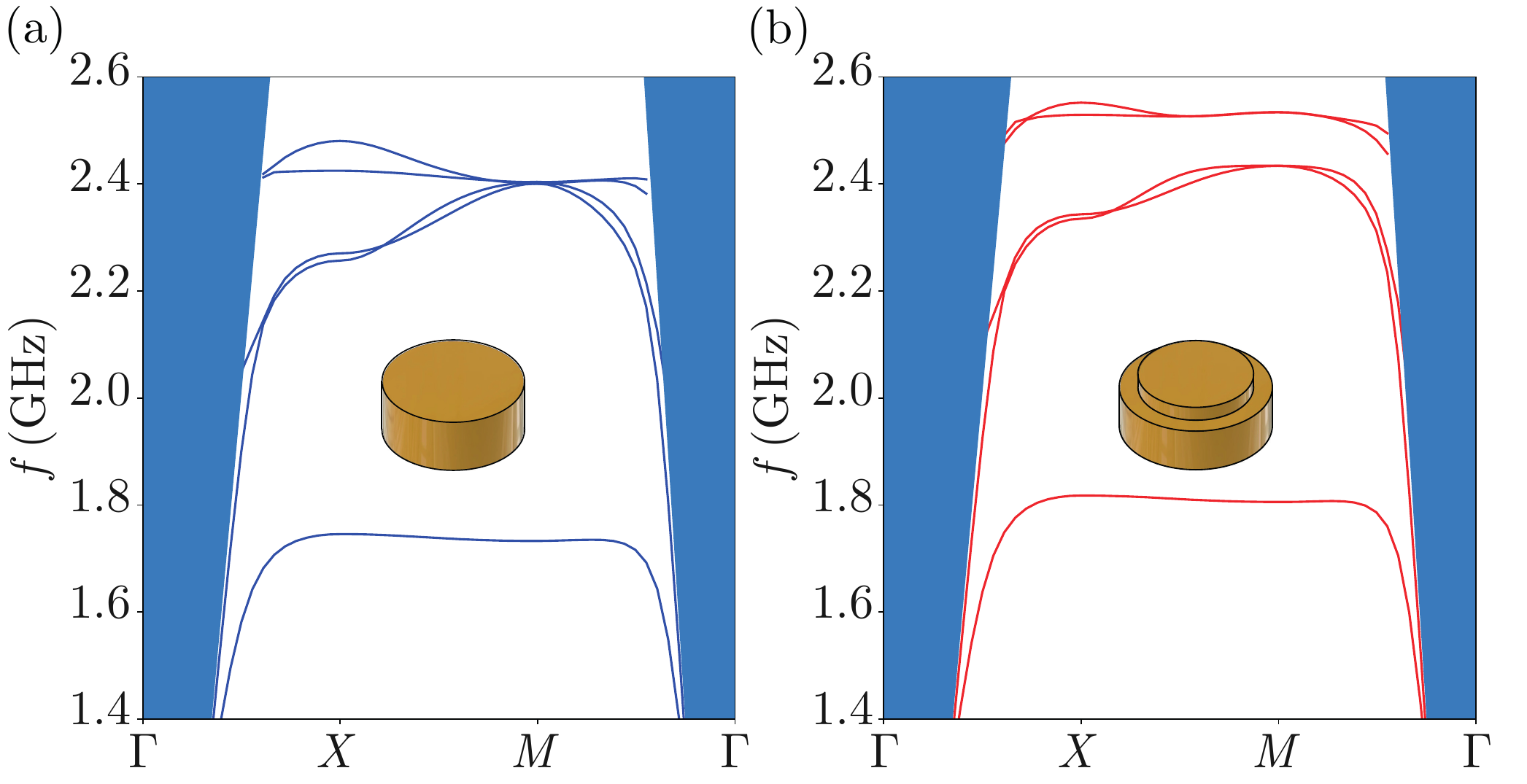}
    \caption{(a) Dispersion diagram for the infinite 2D square lattice consisting of dielectric cylinders. The shaded area shows the light cone, and the inset demonstrates the geometry of a single cylindrical resonator. (b) The same as panel (a), but for the infinite array of bianisotropic resonators.}
    \label{fig:Floquet_dispersion}
\end{figure}

Next, we apply Floquet boundary conditions to the model of a cylindrical resonator composed of ceramics with permittivity $\varepsilon=39$ and having diameter $D = D_{1} = 29.1$~mm and height $h = h_{1} + h_{2} = 12$~mm to simulate an infinite 2D square lattice with period $a=37.1$~mm, which corresponds to the distance $8$~mm between the edges of the resonators. The obtained dispersion diagram shown in Fig.~\ref{fig:Floquet_dispersion}(a) demonstrates a quadratic dispersion in the vicinity of $M$ point along with the absence of a band gap in such a lattice. However, for the same lattice of considered bianisotropic resonators, a band gap ranging from nearly $2.43$ to $2.53$~GHz at the $M$ point is observed; see Fig.~\ref{fig:Floquet_dispersion}(b). This highlights the crucial role of the bianisotropy in the opening of the band gap. The bottom band in the both dispersions corresponds to the magnetic dipole oriented along $z$ axis, which is depicted in Figs.~\ref{fig:CST_Resonator_fields}(b)~and~\ref{fig:CST_Resonator_fields}(c) for the bianisotropic resonator and discussed in the Supplemental Material~\cite{Supplement} for the cylindrical one.

\section{Experimental results}
\label{sec:Experiments}

\subsection{Two-domain configurations}
\label{sec:Experiments_Two_domain}

\begin{figure*}[tbp]
    \includegraphics[width=17.5cm]{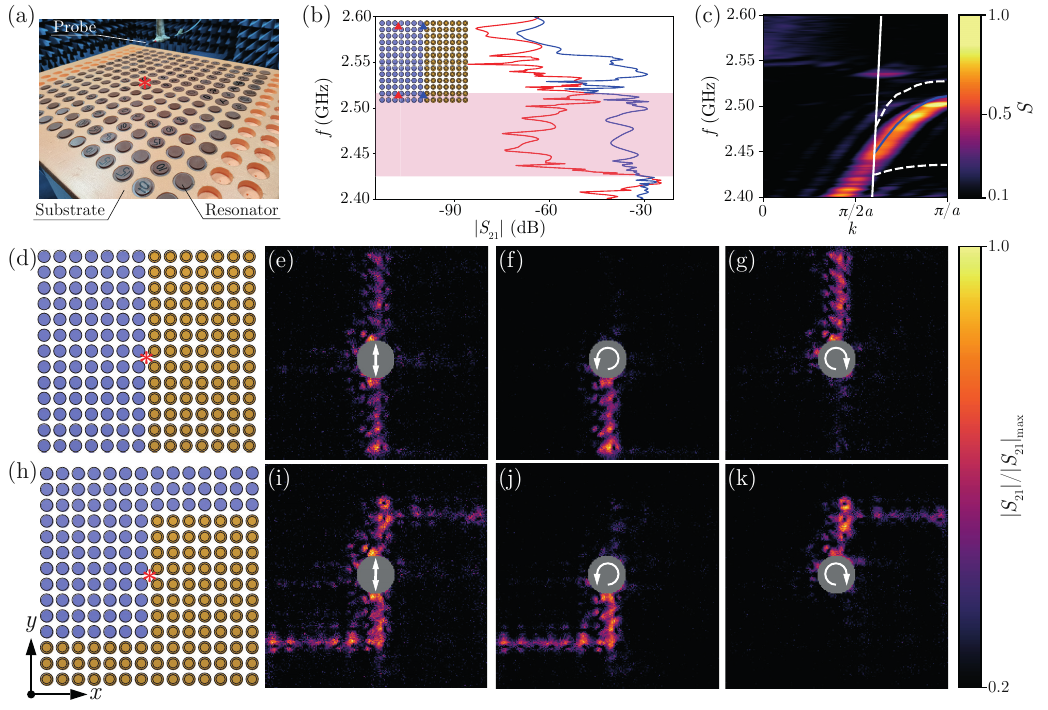}
    \caption{(a) Photo of the considered system implemented as an array of ceramic resonators placed in an extruded polystyrene foam substrate. (b) Transmission between two dipole antennas in the the bulk (red solid line) and at the interface (blue solid line). (c) Experimentally obtained dispersion (dynamic structure factor) for the interface state. The white solid line shows the light cone, while the blue solid line and the white dashed lines demonstrate the edge state dispersion and the band gap edges obtained numerically. (d) Schematics of the configuration with a linear domain wall. (e)-(g) Experimentally measured magnetic fields at $f=2.472$~GHz for the structure with a linear domain wall in the cases of (e) linear, (f) left-handed circular, and (g) right-handed circular polarizations of the excitation. (h)-(k) The same as (d)-(g), but for a double-bend domain wall. The red asterisk in panels (a,d,h) indicates the position of the dipole subwavelength source antenna. The region around the source shaded with gray in panels (e)-(g) and (i)-(k) is not shown for the sake of visual clarity, while the arrows inside this region indicate source polarization in the respective panels.}
    \label{fig:Experimental_results}
\end{figure*}

In the experiments, we consider resonators composed of MgO-CaO-TiO$_2$ ceramics with the same dimensions and permittivity as in the numerical simulations, and arrange them in a square grid of resonators on an extruded polystyrene foam (XPS) substrate with the number of resonators either $13 \times 14$ or $14 \times 14$ in the horizontal plane and the period $a=37.1$~mm, as in Fig.~\ref{fig:Experimental_results}(a). In the following, we start with a structure composed of two oppositely oriented square-lattice arrays of bianisotropic resonators (Fig.~\ref{fig:System}).

First, we characterize the transmission between two subwavelength dipole antennas, the transmitting and the receiving ones, by measuring the $S_{21}$ parameter with a PLANAR~C4490 vector network analyzer (VNA) and an Agilent~83020A microwave amplifier attached to the transmitting antenna. In the first case, the transmitter and the receiver are placed in the bulk of the lattice, the red solid line in Fig.~\ref{fig:Experimental_results}(b), and a considerable decrease in transmission is observed at frequencies between $2.42$ and $2.51$~GHz, facilitating the presence of a band gap. In the second scenario, we place the antennas at the interface between two lattices with opposite orientations of the resonators, represented by the blue solid line in Fig.~\ref{fig:Experimental_results}(b). It is seen that no damping is observed in the band gap region, which highlights the possible excitation of the interface-localized states.

In further experiments, we characterize a near-field distribution of the vertical component of the magnetic field with an XF-R 3-1 H-field probe (a subwavelength loop antenna) while exciting the structure with a subwavelength dipole antenna. Both antennas are attached to a Rohde \& Schwarz ZVB-20 VNA, and measurements are carried out with an automatic Trim TMC~3113 scanner. The scanner step is set to $\Delta=2.5$~mm for both directions in the horizontal plane.

To experimentally characterize the dispersion of interface states, we evaluated the dynamic structure factor (DSF) applying the discrete Fourier transform to the $S_{21}$ parameters extracted along the domain wall, while the source antenna is placed at the edge of the interface; see the Supplemental Material for details~\cite{Supplement}. The normalized DSF obtained in Fig.~\ref{fig:Experimental_results}(c) demonstrates the emergence of the interface-localized excitation in the band gap. Comparison of the extracted dispersion with the light cone $f = ck/2\pi a$, where $c$ is the speed of light, shows that a considerable part of the edge state dispersion curve lies below the light cone. Thus, such edge states should not considerably radiate to free space. Compared to the numerically calculated edge state dispersion [blue solid line in Fig.~\ref{fig:Experimental_results}(c)], the experimentally obtained one is considerably broader. The difference between the measured and numerically evaluated edge state dispersion curves is caused by the finite size of the experimental setup, the deviations in resonant frequencies and positions of the bianisotropic resonators, the presence of noise in the $S_{21}$ parameter measurement, and an imperfect scanning loop positioning. All of the described effects lead to the distortion of periodic patterns in the measured field distributions, which results in the widening of the associated Fourier images. The edges of the band gap extracted from the numerical simulations [white dashed lines in Fig.~\ref{fig:Experimental_results}(c)] are in good agreement with the measured $S_{21}$ parameters in Fig.~\ref{fig:Experimental_results}(b). Details of the numerical calculations of the dispersion curve are given in the Supplemental Material~\cite{Supplement}.

\begin{figure*}[tbp]
    \includegraphics[width=17.5cm]{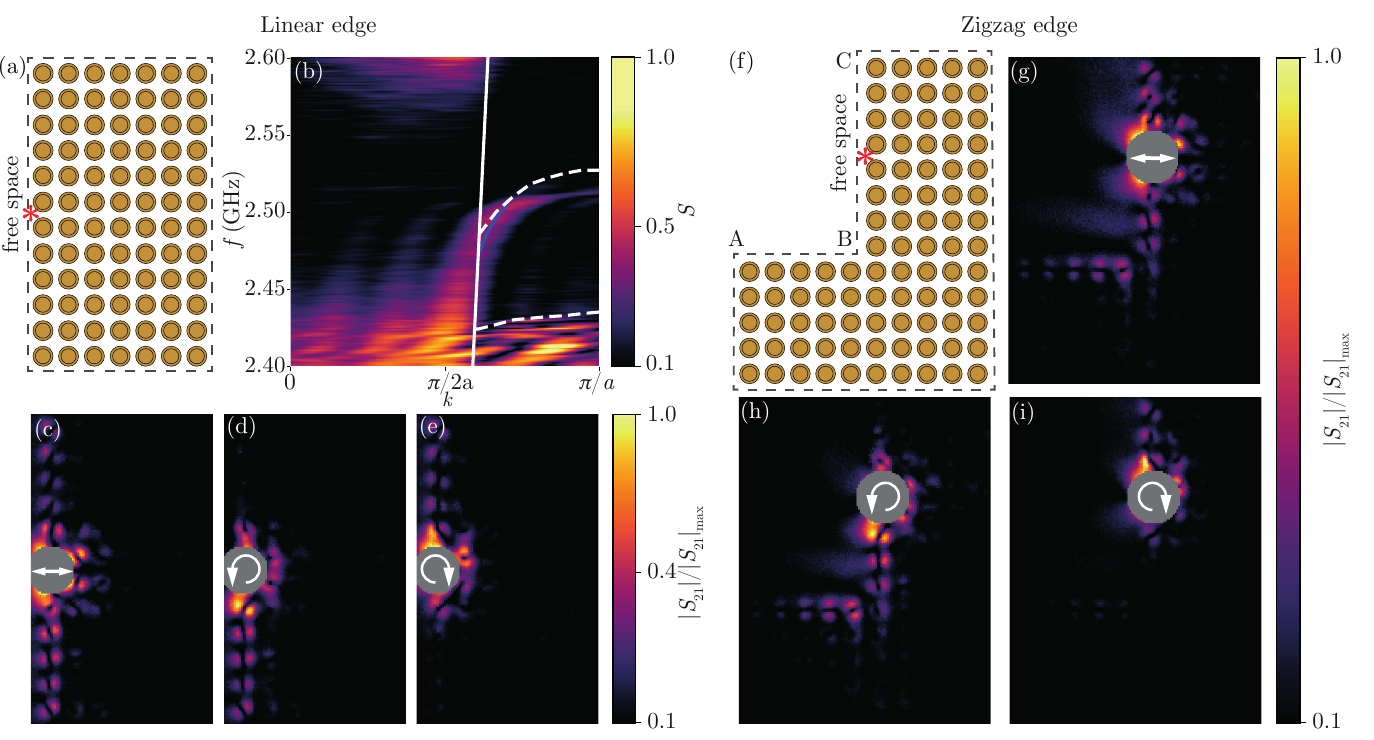}
    \caption{(a) Schematics of the single-domain configuration with a linear boundary. The red asterisk marks the position of the source antenna and the gray dashed line denotes the boundary between the domain with bianisotropic resonators and free space. (b) Experimentally obtained dispersion diagram for the states at the boundary between the domain and free space. The white solid line shows the light cone, while the blue solid line and the white dashed lines demonstrate the edge state dispersion and the bang dap edges obtained numerically. (c)-(e) Magnetic field profiles extracted at the frequency $f = 2.455$~GHz in the case of the sources with (c) linear, (d) left-handed circular, and (e) right-handed circular polarizations. (f) Schematics of the single-domain configuration with a double-bend (zigzag) boundary. The red asterisk marks the position of the source antenna and the gray dashed line corresponds to the boundary between the domain and free space. (g)-(i) Magnetic field profiles extracted at the frequency $f = 2.455$~GHz for sources with (g) linear, (h) left-handed circular, and (i) right-handed circular polarizations.}
    \label{fig:Experiment_single_domain}
\end{figure*}

Next, we perform a near-field scanning of the whole structure with the size $14 \times 13$ cylinders and a linear interface; see Fig.~\ref{fig:Experimental_results}(d). The magnetic field profile obtained for the linear source polarization at the frequency $f = 2.472$~GHz shown in Fig.~\ref{fig:Experimental_results}(e) demonstrates a strong confinement of magnetic fields to the interface in both directions from the transmitting antenna, in agreement with the results in Figs.~\ref{fig:Experimental_results}(b)~and~\ref{fig:Experimental_results}(c).

Next, we test the pseudospin-momentum locking of the observed edge states under a circularly polarized excitation. In such a case, destructive or constructive interference of the fields allows us to achieve unidirectional wave propagation ~\cite{2013_Rodriguez}. To emulate a circularly polarized excitation, we repeat the measurements twice for orthogonal dipole antenna positions and subsequently add or subtract the field profiles with the $\pi/2$ phase shift. It is seen that the considered system supports a unidirectional wave propagation induced by the left-handed [Fig.\ref{fig:Experimental_results}(f)] and right-handed [Fig.\ref{fig:Experimental_results}(g)] circularly polarized sources at frequency $f = 2.472$~GHz, which can be interpreted as the excitation of pseudospin-up and pseudospin-down modes. Such an ability to guide a wave unidirectionally and switch the direction of its propagation is useful from a practical point of view.

To study the robustness of the edge states to sharp bends in the interface geometry, we consider an array of $14 \times 14$ resonators with a double-bend (zigzag) interface, Fig.~\ref{fig:Experimental_results}(h). In analogy to the case with a linear domain wall, we perform a scanning with the linear and circular source polarizations. The distributions obtained at the frequency $f=2.472$~GHz closely resemble those measured for the liner interface configuration, Figs.~\ref{fig:Experimental_results}(i)-\ref{fig:Experimental_results}(k). In particular, a pronounced localization at the interface is observed for the linearly-polarized excitation [Fig.~\ref{fig:Experimental_results}(i)], while the circularly-polarized excitation results in a unidirectional propagation along the interface bend [Figs.~\ref{fig:Experimental_results}(j)~and~\ref{fig:Experimental_results}(k)].

\subsection{Single-domain configurations}
\label{sec:Experiments_Single_domain}

As we demonstrate further, an interesting property of the considered structure is that it also supports edge states at the interface between the resonator array and a free space. Analogous states have been predicted for gyrotropic square-lattice photonic topological insulators~\cite{2021_Tasolamprou} and studied in valley photonic topological insulator systems with the aim of implementing frequency multiplexing in waveguides~\cite{2022_Wei}. Pseudospin-momentum locking at the boundaries of a single-domain topological structure has been demonstrated in 2D photonic crystals with honeycomb lattice, where the band gap opens due to the introduction of shrinking and expanding deformations~\cite{2021_Zhang} or a change in the radii ratio for infinite cylinders in the unit cell~\cite{2020_Chen, 2022_Feng}.

To demonstrate such states, we consider a single domain consisting of $12 \times 7$ resonators, Fig.~\ref{fig:Experiment_single_domain}(a). In a complete analogy to the previously studied interface between the domains with opposite orientations of the resonators in Sec.~\ref{sec:Experiments_Two_domain}, we extract DSF experimentally, compare it with the numerically obtained edge state dispersion [Fig.~\ref{fig:Experiment_single_domain}(b)], and characterize magnetic field distributions in the case of a linearly polarized source [Fig.~\ref{fig:Experiment_single_domain}(c)] and left- or right-handed circular polarizations [Figs.~\ref{fig:Experiment_single_domain}(d),~\ref{fig:Experiment_single_domain}(e)], respectively. Although the localization of the edge state and pseudospin-momentum locking closely resemble those for the interface between two domains, Figs.~\ref{fig:Experimental_results}(e)-\ref{fig:Experimental_results}(g), the edge state dispersion in Fig.~\ref{fig:Experiment_single_domain}(b) is more concentrated above the light cone, facilitating more radiative nature of such states, and demonstrates a nearly flat shape below the light cone, characteristic of lower group velocity compared to the approximately linear behavior of DSF in Fig.~\ref{fig:Experimental_results}(c).

Next, we implement a double-band single-domain configuration, Fig.~\ref{fig:Experiment_single_domain}(f), and characterize its magnetic-field response. When the system is excited by a subwavelength dipole antenna, the magnetic field concentrated at the cylinders nearest to the free space region is able to envelope the angle with vertex $B$, which is not observed for the angle with vertex $C$. Moreover, the field drastically fades before reaching the vertex $A$, Fig.~\ref{fig:Experiment_single_domain}(g). The same effect manifests for the system excited by left- or right-handed circularly polarized sources, and the one-way propagation is also observed; see Figs.~\ref{fig:Experiment_single_domain}(h)~and~\ref{fig:Experiment_single_domain}(i). In the Supplemental Material~\cite{Supplement}, we also provide experimental results for disordered geometries with extracted resonators at the interface, which highlight that edge states lack robustness towards such point defects.

\begin{figure}[tbp]
    \includegraphics[width=8.5cm]{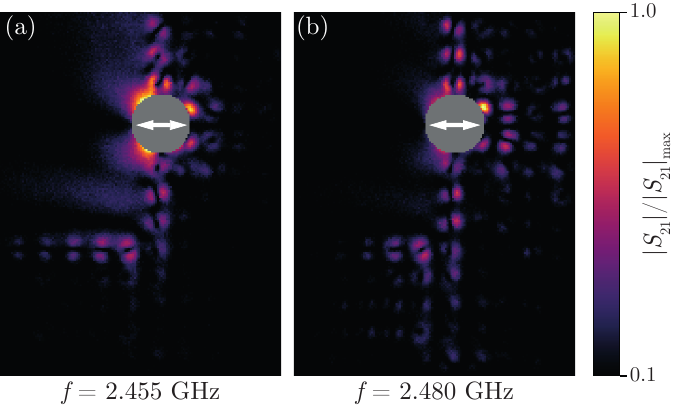}
    \caption{Experimentally obtained magnetic field profiles for the single-domain configuration with the zigzag edge excited by a linearly polarized source at the frequencies (a) $f = 2.455$~GHz and (b) {$f = 2.480$}~GHz.}
    \label{fig:Radiation}
\end{figure}

Finally, it is worth noting that a considerable portion of edge state dispersion both for the two-domain configuration in Fig.~\ref{fig:Experimental_results}(c) and especially for a single-domain configuration in Fig.~\ref{fig:Experiment_single_domain}(b) is above the light cone. Indeed, it is seen that the experimentally obtained magnetic field profiles at the frequency $f = 2.455$~GHz possess energy radiation in free space [Fig.~\ref{fig:Radiation}(a)], while the fields at the frequency $f = 2.480$~GHz localize inside the structure and demonstrate more pronounced coupling with bulk excitations, as the frequency of such a state is close to the bulk spectrum [Fig.~\ref{fig:Radiation}(b)].

\section{Theoretical model and topological properties of the system}
\label{sec:Hamiltonian}

According to the electric and magnetic field distributions of an individual bianisotropic resonator, see Fig.~\ref{fig:CST_Resonator_fields} and Supplemental Material Note~1~\cite{Supplement}, in the frequency range from $2.39$ to $2.53$~GHz the system can be considered as an array of coupled electric and magnetic point dipoles located and directed in the $(xy)$ plane. To devise the theoretical model, we rely on the dyadic Green's function approach outlined for a one-dimensional chain of bianisotropic resonators in Ref.~\cite{2019_Gorlach}. We proceed with taking into account the interactions between the nearest resonators separated by the lattice constant $a$ and diagonally opposite next-nearest resonators separated by the distance $\sqrt{2}a$.

\begin{figure*}[tbp]
    \includegraphics[width=17.5cm]{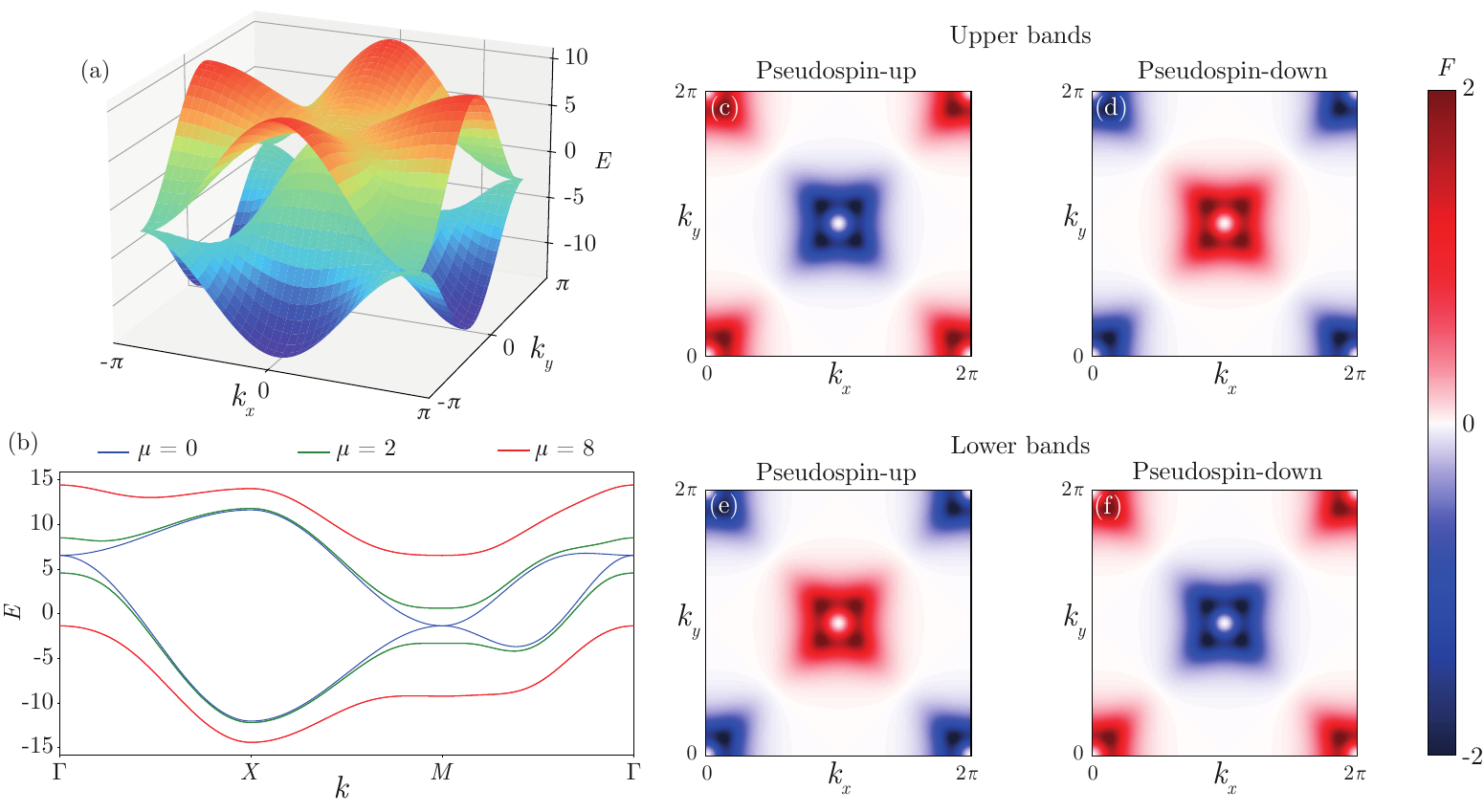}
    \caption{(a) Eigenenergy bands, Eq.~\eqref{eq:E_k}, in the case of absent bianisotropy $\mu = 0$. Color shows the value of $E(k)$. (b) The dispersion diagrams for the eigenvalues $E(k)$ of the Hamiltonian~\eqref{eq:E_k} for the wave number $k=(k_{x},k_{y})$ varied along the $\Gamma-X-M-\Gamma$ trajectory in the reciprocal space for different values of the bianisotropy, $\mu=0$ (blue solid line), ${\mu=2}$ (green solid line), and $\mu=8$ (red solid line). (c)-(f) The distributions of Berry curvature $F(k_{x},k_{y})$ in the Brillouin zone for $\mu=1$ evaluated at $101 \times 101$ equidistantly distributed points $(k_{x},k_{y})$. Color shows the value of the Berry curvature.}
    \label{fig:Hamiltonian}
\end{figure*}

Within such an approach, each resonator is characterized by a four-component vector of its in-plane electric $p_{x,y}$ and magnetic $m_{x,y}$ moments that can grouped into a pseudospin basis $\ket{\psi} = (p_{x}+m_{x}, p_{y}+m_{y}, p_{x}-m_{x}, p_{y}-m_{y})^{\rm T}$~\cite{2016_Slobozhanyuk, 2017_Slobozhanyuk}. In this basis, the effective Bloch Hamiltonian $\hat{H}(k_{x},k_{y})$ for the infinite 2D square lattice array of resonators takes the following form:
\begin{multline}
    \hat{H}(k_{x},k_{y}) = \big(\sqrt{2} \cos k_{x} \cos k_{y} + 2 \cos k_{x} + 2 \cos k_{y} \big)\hat{I}\otimes\hat{I} + \\
    + 6(\cos k_{x} - \cos k_{y}) \hat{I}\otimes\hat{\sigma}_{z} - 3\sqrt{2} \sin k_{x} \sin k_{y} \hat{I}\otimes\hat{\sigma_x} - \\ 
    - \mu \hat{\sigma}_z\otimes\hat{\sigma}_y =  
   \begin{pmatrix}
       \hat{H}^{\uparrow} & 0\\
       0 & \hat{H}^{\downarrow}
   \end{pmatrix},
    \label{eq:Bloch_Hamiltonian}
\end{multline}
where $\hat{I}$ is the $[2 \times 2]$ unity matrix, $\hat{\sigma_{x}} = (0, 1; 1, 0)$, $\hat{\sigma_{y}} = (0, -{\rm i}; {\rm i}, 0)$, and $\hat{\sigma_{z}}=(1, 0; 0, -1)$ are Pauli matrices, the parameter $\mu$ characterizes the strength of the bianisotropic coupling of electric and magnetic moments, and $k_{x,y}$ are the in-plane components of the wave number. The Bloch Hamiltonian derivation is detailed in Supplemental Material Note~4~\cite{Supplement}. The diagonal $[2 \times 2]$ blocks $\hat{H}^{\uparrow}$ and $\hat{H}^{\downarrow}$ describe two Kramers partners analogous to pseudospin-up and pseudospin-down states in the selected basis and share the same set of eigenvalues. Such a Hamiltonian thus features two doubly degenerate bands of eigenenergies $\hat{H}(k_{x},k_{y})\ket{\psi} = E(k_{x},k_{y})\ket{\psi}$:
\begin{multline}
    E^{\uparrow(\downarrow)}(k_x, k_y) = 2\big(\cos k_x + \cos k_y\big) + \sqrt{2} \cos k_x \cos k_y \pm \\
     \pm \frac{3}{\sqrt{2}}\big(9 + \frac{2}{9}\mu^2 - 16 \cos k_x \cos k_y + 3\cos(2k_y) +\\
     + 3\cos(2k_x) + \cos(2k_x)\cos(2k_y)\big)^{1/2},
    \label{eq:E_k}
\end{multline}
which are visualized for arrays without bianisotropy ($\mu=0$) in Fig.~\ref{fig:Hamiltonian}(a). Also, it is seen from Eq.~\eqref{eq:E_k} that at high-symmetry points of the Brillouin zone, $\Gamma$ ($k_{x} = k_{y} = 0$) and $M$ ($k_{x} = \pm k_{y} = \pm \pi$), all bands become fourfold degenerate, and the band gap quadratically closes; see Fig.~\ref{fig:Hamiltonian}(b). However, adding a bianisotropy $\mu > 0$ eliminates such a degeneracy, resulting in a band gap opening for every $k$, see the example for $\mu=2$ in Fig.~\ref{fig:Hamiltonian}(b). Finally, for sufficiently large $\mu$, a complete band gap is observed, shown in Fig.~\ref{fig:Hamiltonian}(b) for $\mu=8$. Eigenfunctions of the Hamiltonian~\eqref{eq:Bloch_Hamiltonian} in the explicit form are given in the Supplemental Material~\cite{Supplement}.

Considering previously demonstrated pseudospin-locking, robust propagation of the in-gap edge states in experiments, and the form of Bloch Hamiltonian~\eqref{eq:Bloch_Hamiltonian}, the question naturally arises if the emerging band gap is topological, and the considered system supports photonic quantum spin-Hall phase similar to the ones reported in Refs.~\cite{2012_Khanikaev, 2017_Slobozhanyuk, 2019_Slobozhanyuk}. Since the considered system possesses time-reversal symmetry, the Chern numbers of the upper and lower bands vanish~\cite{2007_Fu}. Moreover, the breaking of parity symmetry along the $z$ axis zeros out the integral of Berry curvature over the Brillouin zone, and local Berry curvature in the vicinity of $\Gamma$ and $M$ high-symmetry points should be considered instead of evaluating spin Chern numbers~\cite{2020_Jin, 2024_Zou}.

To calculate the Berry curvature $F(k_x, k_y)$, one needs to evaluate the following expression for each of the eigenfunctions $\ket{\psi_{u(l)}^{\uparrow(\downarrow)}}$ of the Hamiltonian $\hat{H}^{\uparrow(\downarrow)}$, Eq.~\eqref{eq:Bloch_Hamiltonian}, in the pseudospin basis defined above:
\begin{multline}
    F(k_x, k_y) = \dfrac{\partial}{\partial k_x}\braket{\psi_{u(l)}^{\uparrow(\downarrow)} | \dfrac{\partial}{\partial k_y}| \psi_{u(l)}^{\uparrow(\downarrow)}} -\\
    - \dfrac{\partial}{\partial k_y}\braket{\psi_{u(l)}^{\uparrow(\downarrow)} | \dfrac{\partial}{\partial k_x}| \psi_{u(l)}^{\uparrow(\downarrow)}},
    \label{eq:Berry_Curvature}
\end{multline}
where the superscripts $\uparrow$ and $\downarrow$ indicate the pseudospin-up ($\hat{H}^{\uparrow}$) and the pseudospin-down ($\hat{H}^{\downarrow}$) parts of the Hamiltonian~\eqref{eq:Bloch_Hamiltonian}, and the subscripts $u$ and $l$ denote the upper and lower bands of $\hat{H}^{\uparrow (\downarrow)}$, respectively. The Berry curvature distributions obtained for each of the eigenfunctions for the parameter $\mu \ne 0$ are shown in Figs.~\ref{fig:Hamiltonian}(c)-\ref{fig:Hamiltonian}(f). It is seen that the pseudospin-up and pseudospin-down parts of the Hamiltonian feature equivalent distributions of the Berry curvature, but with opposite signs for the upper [Figs.~\ref{fig:Hamiltonian}(c)~and~\ref{fig:Hamiltonian}(d)] and the lower [Fig.~\ref{fig:Hamiltonian}(e)~and~\ref{fig:Hamiltonian}(f)] bands, which demonstrates the topological origin of the observed edge states~\cite{2020_Jin, 2024_Zou}. In the absence of bianisotropy, the Berry curvature zeros out throughout the Brillouin zone for all four eigenfunctions, indicating the topologically trivial nature of excitations in such systems, as can be verified by a direct calculation of Eq.~\eqref{eq:Berry_Curvature} for eigenfunctions defined in Supplemental Material Note~4~\cite{Supplement} considering $\mu=0$. Thus, the introduction of bianisotropy is crucial for changing the topology of the 2D square lattice resonator array from a trivial to a nontrivial one. It is important to note that in the nearest-neighbor approximation (when only the four nearest resonators separated from the considered one by the lattice constant $a$ are taken into account), the band gap readily emerges for $\mu > 0$, but Berry curvature still vanishes throughout the Brillouin zone for arbitrary values of the parameter $\mu$, as can be checked by a direct evaluation of Eq.~\eqref{eq:Berry_Curvature} for the eigenfunctions in the nearest-neighbor approximation; see Supplemental Material Note~4~\cite{Supplement} for details. Thus, such a model is insufficient for a correct description of topological properties of the considered system, facilitating the need to take into account long-range interactions between the resonators; at least the couplings between the next-nearest ones. An important role for couplings beyond the nearest-neighbor approximation has been demonstrated considering the emergence of a band gap that hosts corner-localized states in the 2D Su-Schrieffer-Heeger model~\cite{2022_D4} and the formation of the second type of corner states governed by long-range interactions in the 2D kagome lattice of coupled resonators~\cite{2020_Li}. Similar diagonal couplings also arise in certain spatial regions when considering equivalent 2D tight-binding models describing topological phenomena in 1D two-particle quantum systems with nonlinear interactions~\cite{2020_Olekhno, 2022_Olekhno}.

\section{Discussion}
\label{sec:Discussion}

To conclude, we have numerically and experimentally demonstrated the emergence of edge states at microwave frequencies supported by a simple square lattice of resonators and governed by their individual bianisotropy. The demonstrated phenomena include pseudospin-momentum locking of such states at the interface between two distinct domains and their resilience towards geometrical imperfections. All of the reported effects are also observed at the boundary between a single domain and free space, up to a slight modification of the edge-state dispersion. The considered system possesses nonzero local Berry curvature, and thus represents a photonic analog of a quantum spin Hall insulator.

As directions for further development, thermal~\cite{2022_Kurganov} or mechanical~\cite{2022_Xu} reconfigurability may be introduced to the structures considered to engineer such edge states. The other prospective avenue is to consider three-phase arrays composed of resonator domains with different signs of bianisotropy as well as free space or domains consisting of resonators without pronounced bianisotropy. Such systems have been considered in Ref.~\cite{2023_Li} for realizing on-chip THz waveguides. It is also interesting to address nonlinear effects~\cite{2018_Mittal, 2019_DAguanno, 2019_Kruk, 2020_Smirnova, 2024_Sone} and quantify the effects of long-range interactions~\cite{2020_Li, 2021_Vakulenko} in the model considered. Finally, the observed phenomena can be realized in the optical domain by miniaturizing the resonators~\cite{2021_Zhirihin}.

\section*{Acknowledgments}

We acknowledge fruitful discussions with Maxim Gorlach. The work was financially supported by the Russian Science Foundation (Project No.~24-72-10069).

\end{document}


\title{Supplemental Material\\
Topological edge states in all-dielectric square-lattice arrays of bianisotropic microwave resonators}

\author{Alina~D.~Rozenblit}
  \email{alina.rozenblit@metalab.ifmo.ru}
\affiliation{School of Physics and Engineering, ITMO University, 49 Kronverksky pr., bldg. A, 197101 Saint Petersburg, Russia}

\author{Georgiy~D.~Kurganov}
\affiliation{School of Physics and Engineering, ITMO University, 49 Kronverksky pr., bldg. A, 197101 Saint Petersburg, Russia}

\author{Dmitry~V.~Zhirihin}
\affiliation{School of Physics and Engineering, ITMO University, 49 Kronverksky pr., bldg. A, 197101 Saint Petersburg, Russia}

\author{Nikita~A.~Olekhno}
\affiliation{School of Physics and Engineering, ITMO University, 49 Kronverksky pr., bldg. A, 197101 Saint Petersburg, Russia}

\date{\today}

\maketitle

\tableofcontents

\section*{Supplementary Note 1. Modes of individual resonators}

\begin{figure*}[tbp]
    \includegraphics[width=10cm]{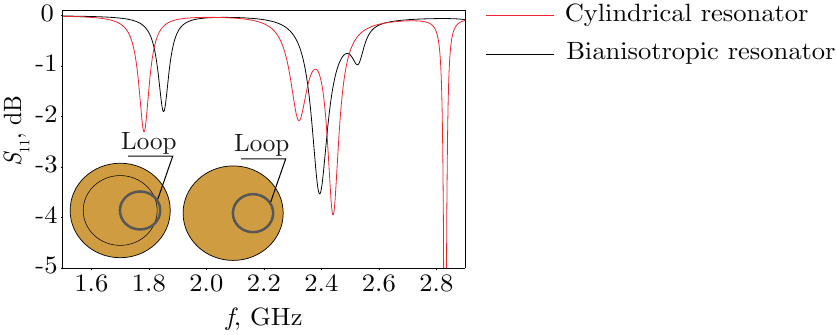}
    \caption{Numerically calculated $S_{11}$ parameters for the cylindrical (red solid line) and the bianisotropic (black solid line) resonators. The inset show the position of the loop antenna for both types of resonators.}
    \label{fig:S11_bianisotropic_and_cylindrical}
\end{figure*}

To study the properties of the individual resonator, we numerically evaluate the resonant response of the cylindrical and bianisotropic resonators using CST Microwave Studio. In particular, we calculate numerically $S_{11}$ parameter by introducing a loop antenna with diameter $12$~mm implemented as a perfect electric conductor wire with thickness $0.4$~mm which is placed atop the resonator at a height of $1$~mm. The obtained spectra are shown in Fig.~\ref{fig:S11_bianisotropic_and_cylindrical}. Next, to study the field distributions of the resonances, we perform numerical simulations using the eigenmode Solver for cylindrical and bianisotropic resonators surrounded by electric boundaries set at $15$~cm away from the edges of the resonator. To visualize the field distributions obtained, we plot the components of the magnetic field ($H_x$, $H_y$, $H_z$) and the electric field ($E_x$, $E_y$, $E_z$) for three orthogonal planes crossing the center of the resonator.

\begin{figure*}[tbp]
    \includegraphics[width=16cm]{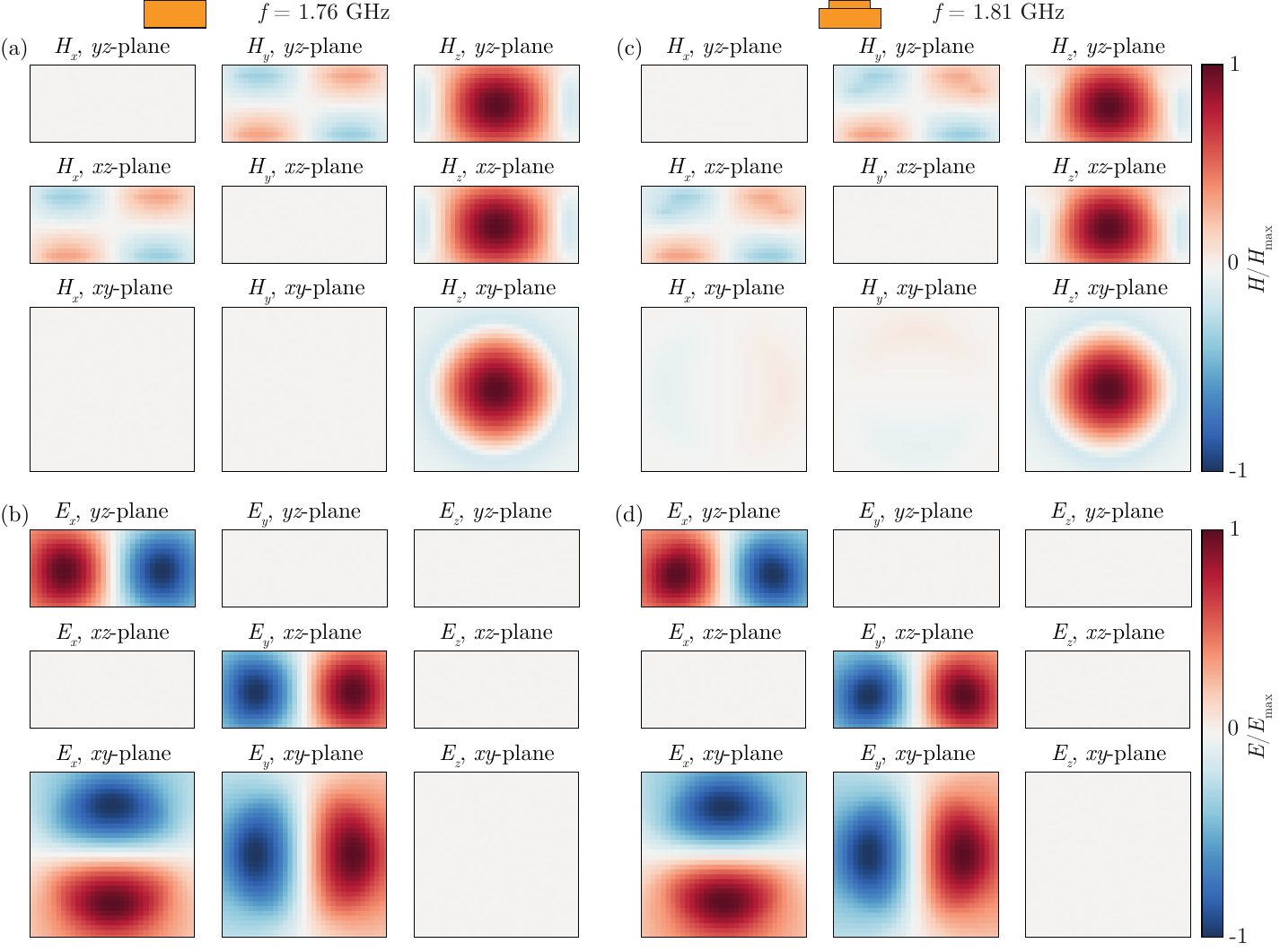}
    \caption{Magnetic field $H$ and electric field $E$ components extracted for three orthogonal planes crossing the center of the resonator for (a),(b) the cylindrical resonator at the frequency $f = 1.76$~GHz and (c),(d) the bianisotropic resonator at the frequency $f = 1.81$~GHz. Magnetic and electric field values (shown with color) are normalized to the maximal value of the corresponding field amplitude among all components and planes in each panel. The image plane orientations $(xy)$, $(xz)$, and $(yz)$ are specified above each subplot.}
    \label{fig:Single_resonator_peak_1}
\end{figure*}

The distributions of field components corresponding to the first resonant peak of the cylindrical ($f = 1.76$~GHz) and the bianisotropic ($f = 1.81$~GHz) resonators are shown in Fig.~\ref{fig:Single_resonator_peak_1}. As seen from the extracted field profiles, in both cases the first resonant peak is associated with the magnetic dipole oriented along the $z$ axis. Indeed, Figs.~\ref{fig:Single_resonator_peak_1}(a),(c) demonstrate a nearly uniform magnetic field at the symmetry axis of the resonator which is directed vertically, while Figs.~\ref{fig:Single_resonator_peak_1}(b),(d) clearly correspond to a toroidal electric field distribution in the $(xy)$ plane which creates a circular current manifesting as such a magnetic dipole. The frequency of the mode increases for the bianisotropic resonator due to the decrease from $12$~mm to $9$~mm in the height of the disk with a greater diameter.

\begin{figure*}[tbp]
    \includegraphics[width=16cm]{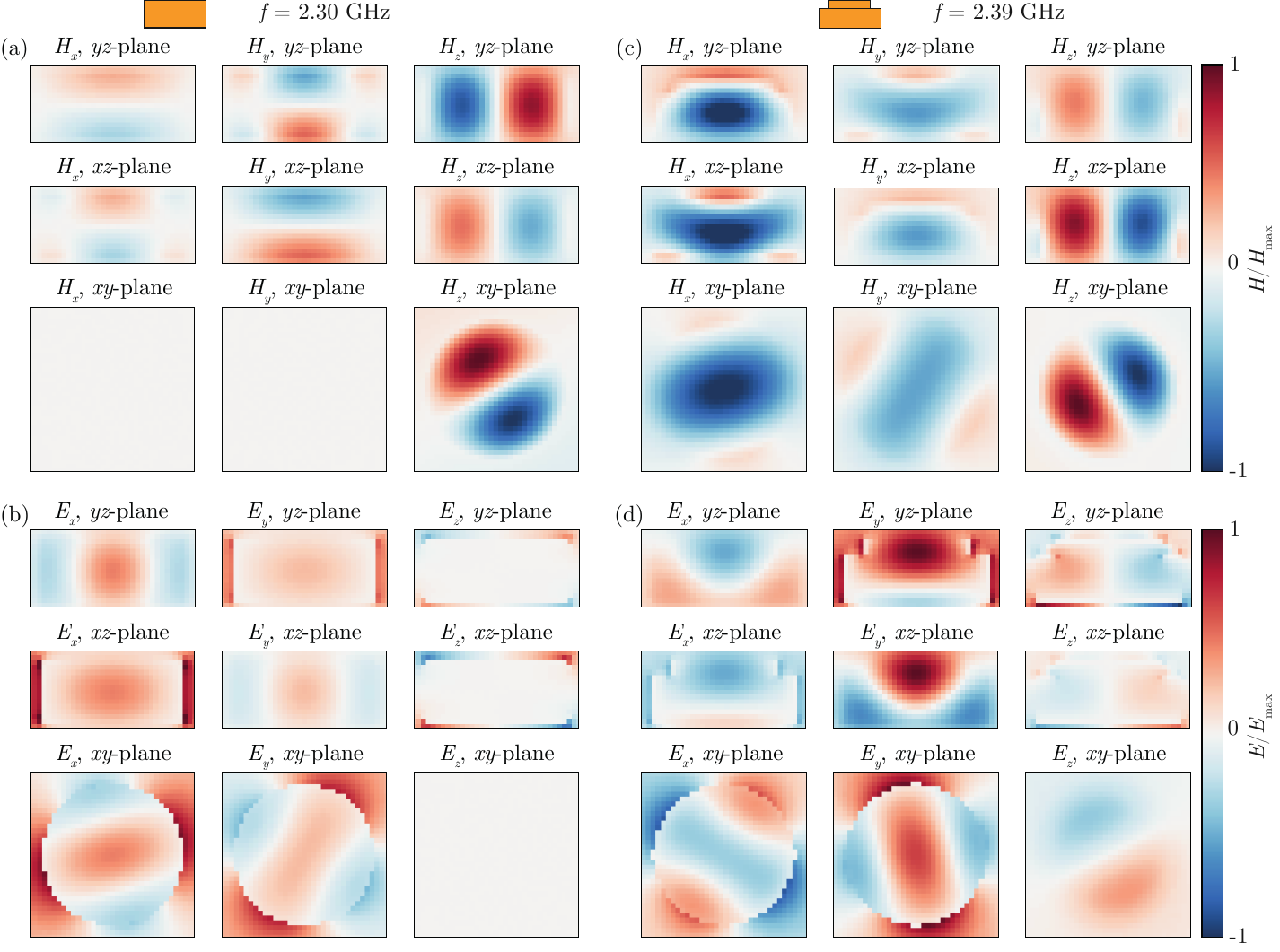}
    \caption{Magnetic field $H$ and electric field $E$ components extracted for three orthogonal planes crossing the center of the resonator for (a),(b) the cylindrical resonator at the frequency $f = 2.30$~GHz and (c),(d) the bianisotropic resonator at the frequency $f = 2.39$~GHz. Magnetic and electric field values (shown with color) are normalized to the maximal value of the corresponding field amplitude among all components and planes in each panel. The image plane orientations $(xy)$, $(xz)$, and $(yz)$ are specified above each subplot.}
    \label{fig:Single_resonator_peak_2}
\end{figure*}

In turn, the field components distribution corresponding to the second resonant peak of the cylindrical resonator at the frequency $f = 2.30$~GHz represents the electric dipole in the $(xy)$ plane, as seen in Figs.~\ref{fig:Single_resonator_peak_2}(a),(b). In particular, it is seen that the magnetic field in Fig.~\ref{fig:Single_resonator_peak_2}(a) demonstrates a toroidal shape similar to a field created by a wire with current located in the $(xy)$ plane, which is consistent with the electric field distributions in Fig.~\ref{fig:Single_resonator_peak_2}(b) facilitating vanishing $z$ component of the electric field and a nearly uniform in-plane electric field concentrated at the diameter of the resonator. Thus, this mode represents an electric dipole.

In contrast to the case of the cylindrical resonator, the field profiles for the bianisotropic resonator in Figs.~\ref{fig:Single_resonator_peak_2}(c),(d) demonstrate the simultaneous presence of features characteristic for electric and magnetic dipoles. Indeed, the $z$-component of the magnetic field in Fig.~\ref{fig:Single_resonator_peak_2}(c) demonstrates a toroidal magnetic current created by an electric dipole in the $(xy)$ plane, while, at the same time, the $x$ and $y$ components of the magnetic field have the typical shape of a magnetic dipole in the $(xy)$ plane. However, these dipole patterns appear to be distorted compared to those of Figs.~\ref{fig:Single_resonator_peak_1}(c),(d), indicating the formation of a hybrid mode. Also, it is seen that the electric and magnetic dipole moments within such a mode are orthogonal. The same picture is supported by the analysis of the electric field components in Fig.~\ref{fig:Single_resonator_peak_2}(d), which demonstrate the presence of electric dipole features seen in the $E_{x}$ and $E_{y}$ components in the $(xy)$ plane that are accompanied by toroidal currents in the $(xy)$ and $(yz)$ planes. The analyzed field distributions thus indicate that the presence of bianisotropy leads to the formation of hybrid modes from the electric and magnetic dipole resonances in the $(xy)$ plane.

\begin{figure*}[tbp]
    \includegraphics[width=16cm]{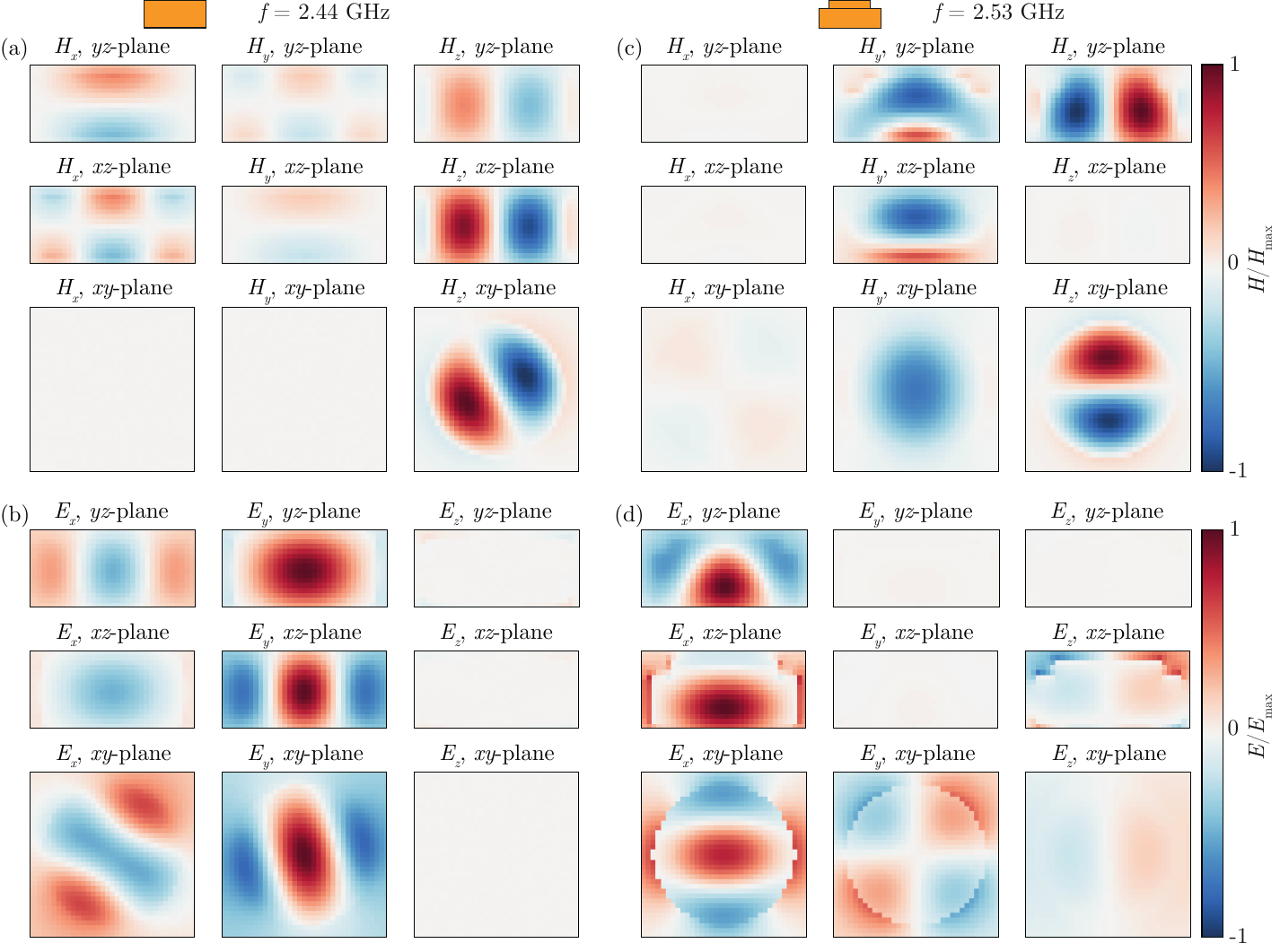}
    \caption{Magnetic field $H$ and electric field $E$ components extracted for three orthogonal planes crossing the center of the resonator for (a),(b) the cylindrical resonator at the frequency $f = 1.44$~GHz and (c),(d) the bianisotropic resonator at the frequency $f = 1.53$~GHz. Magnetic and electric field values (shown with color) are normalized to the maximal value of the corresponding field amplitude among all components and planes in each panel. The image plane orientations $(xy)$, $(xz)$, and $(yz)$ are specified above each subplot.}
    \label{fig:Single_resonator_peak_3}
\end{figure*}

For the third resonance of the cylindrical resonator at $f=2.44$~GHz, the field distributions in Figs.~\ref{fig:Single_resonator_peak_3}(a),(b) closely resemble those of Figs.~\ref{fig:Single_resonator_peak_2}(a),(b) and correspond to the electric dipole located in the $(xy)$ plane and orthogonal to the electric dipole at frequency $f = 2.30$~GHz. In the bianisotropic resonator, the third resonance at the frequency $f=2.53$~GHz shown in Figs.~\ref{fig:Single_resonator_peak_3}(c),(d) demonstrates features of the electric dipole oriented along the $x$ axis and the magnetic dipole oriented along the $y$ axis, as seen from the distributions $H_{z}$ and $H_{y}$ in Fig.~\ref{fig:Single_resonator_peak_3}(c), respectively. The same is seen from the $x$- and $z$-components of the electric field in Fig.~\ref{fig:Single_resonator_peak_3}(d). Thus, the third resonance in the bianisotropic resonator also corresponds to a hybrid mode.

\section*{Supplementary Note 2. Numerical model and simulation results}

\textit{Unit cell simulations.} For simulations of the infinite array of bianisotropic resonators, we use the eigenvalue solver applied to the unit cell under Floquet boundary conditions along the $x$ and $y$ axes, while both $k_x$ and $k_y$ wavevectors change along the $\Gamma - X - M - \Gamma$ trajectory in the reciprocal space between the high-symmetry points of the Brillouin zone. The dispersion diagrams obtained for the infinite array of bianisotropic and cylindrical resonators with period $a = 37.1$~mm are shown in Figs.~1(a),(b) in the main text.

\begin{figure*}[tbp]
    \includegraphics[width=15cm]{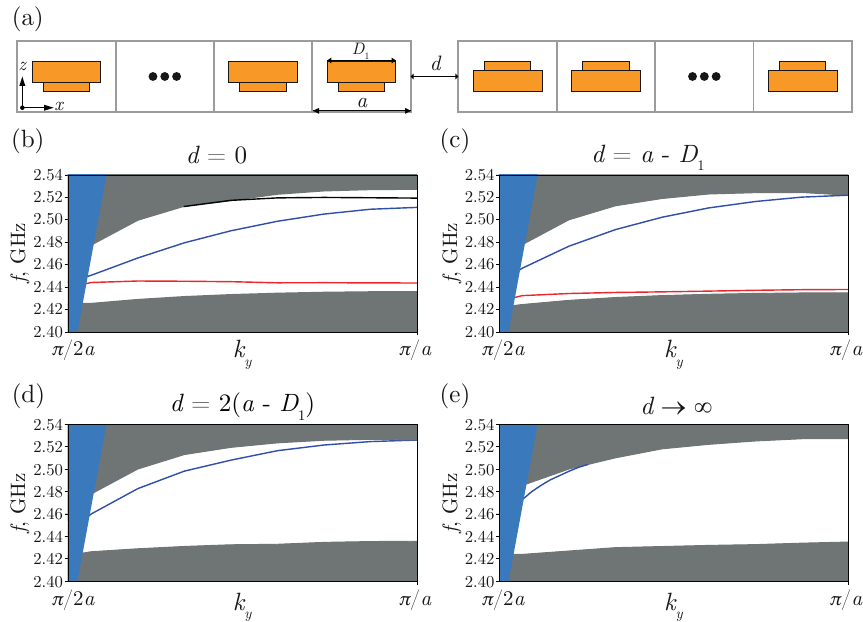}
    \caption{(a) Schematics of the supercell consisting of two oppositely oriented domains separated by the shift $d$, where $a=37.1$~mm and $D_{1}=29.1$~mm indicate the lattice constant and the greater diameter of the resonator, respectively. (b)-(d) Dispersion diagrams for supercells with different shifts $d$ between the domains with opposite signs of bianisotropy for (b) $d = 0$, (c) $d = a-D_{1}$, and (d) $d = 2(a-D_{1})$. (e) Dispersion diagram for the supercell consisting of a single domain and vacuum media replacing the domain with the opposite orientation of resonators, which is equivalent to the infinite separation between the domains $d \to \infty$. The gray and blue shaded areas correspond to the bulk bands and radiation continuum above the light cone, respectively. The black, blue, and red solid lines indicate the modes with electromagnetic fields localized at the interfaces between two domains or between the domain and vacuum media.}
    \label{fig:CST_Supercell}
\end{figure*}

\textit{Supercell simulations.} The supercell numerical model includes two 1D domains with opposite orientations of resonators placed along the $x$ axis and separated by a spacing $d$ between the domains. Each domain includes seven resonators. In this case, the Floquet boundary conditions are also set in both $x$ and $y$ directions. However, the wave number along the $x$ axis of the supercell is set to $k_x=0$ corresponding to periodic boundary conditions, while the wave number $k_y$ varies from $0$ to $\pi/a$. Note that such an approach leads to the emergence of an additional boundary interface associated with periodic boundary conditions. In the following, we select only the modes corresponding to the central interface, increase the distance $d$ between the domains, and finally replace one of them with a vacuum medium. The latter model allows us to search for the modes that feature energy localization along the interface between the domain and the free space.

The dispersion diagram obtained for $d=0$ shown in Fig.~\ref{fig:CST_Supercell}(b) demonstrates the existence of three isolated curves corresponding to the in-gap edge states, with two of them (the black and red curves) being close to the bulk bands. When the distance between the domains increases to $d = a-D_{1}$, only two types of states are observed, as one of the initial states merges into the upper bulk band, Fig.~\ref{fig:CST_Supercell}(c). For $d = 2(a-D_{1})$, only a single type of edge state exists, as the red branch merges into the lower bulk continuum, Fig.~\ref{fig:CST_Supercell}(d). At the same time, the states on the blue branch corresponding to the experimentally observed states [Fig.~4(c) in the main text] shift upward with increasing spacing between the domains, and, finally, in the supercell with a single domain, just a slight tail of these modes remains below the light cone, Fig.~\ref{fig:CST_Supercell}(e). In the experiments, we do not observe the states corresponding to the black and red branches because of their hybridization with the bulk modes.

\begin{figure*}[tbp]
   \includegraphics[width=18cm]{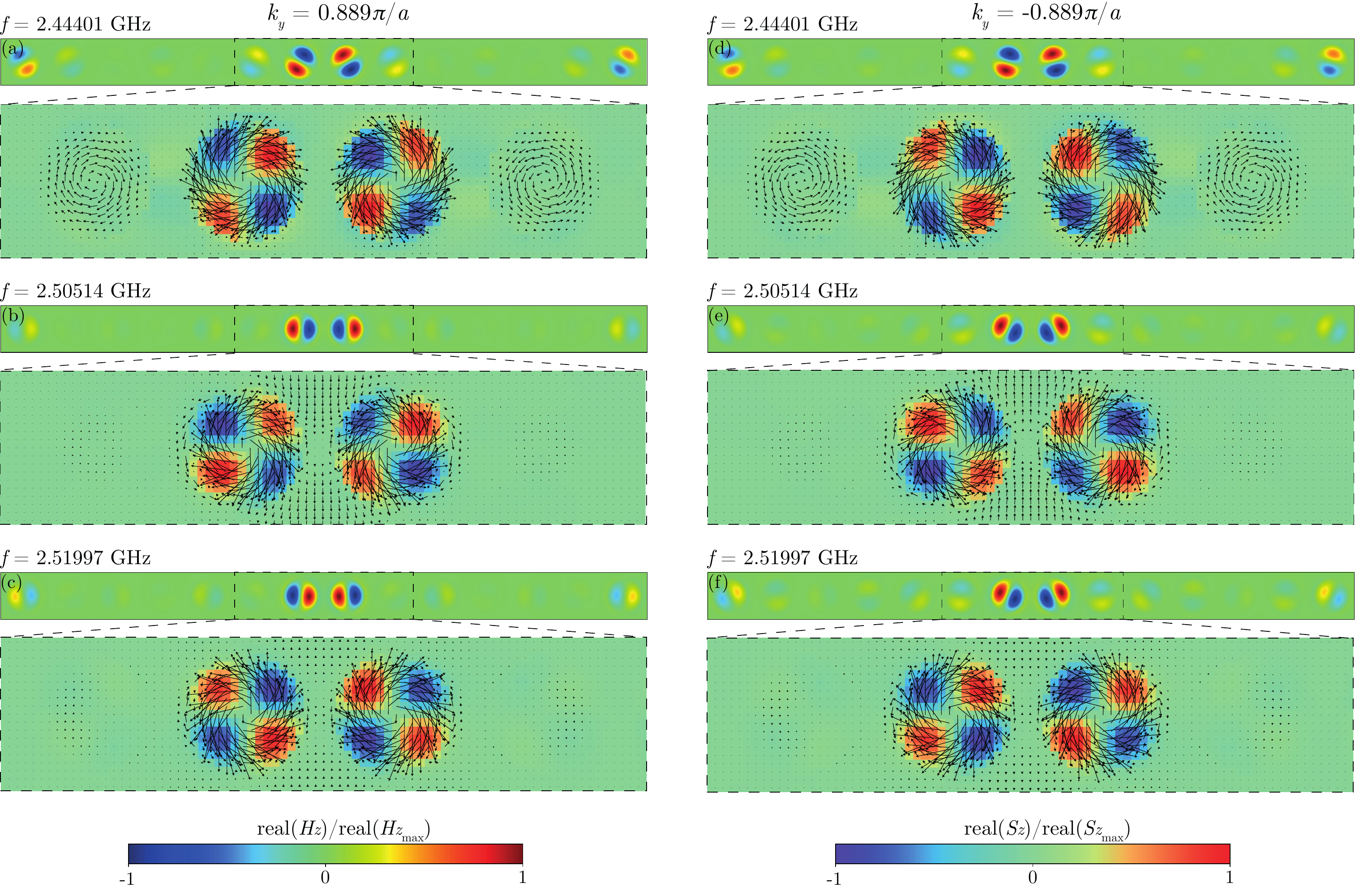}
   \caption{Numerically obtained distributions of the $H_z$ magnetic field component. The insets demonstrate the normal (shown by color) and tangential (shown by black arrows) Poynting vector components for the zoomed-in interface region. The results are calculated for the supercell consisting of two domains with opposite orientations of resonators for wave numbers (a)-(c) $k_y = 0.889 \pi /a$ and (d)-(f) $k_y = -0.889 \pi /a$. The panels correspond to the different frequencies in the bandgap: (a),(d) $f=2.44401$~GHz; (b),(e) $f=2.50514$~GHz; and (c),(f) $f=2.51997$~GHz.}
   \label{fig:Supercell_profiles}
\end{figure*}

Figure~\ref{fig:Supercell_profiles} demonstrates the normal magnetic field component $H_z$ distributions and the Poynting vector for zoomed-in regions of the supercell ($d=0$) under the following boundary conditions: $k_y = 0.889 \pi /a$ [Figs.~\ref{fig:Supercell_profiles}(a)-(f)] and $k_y = -0.889 \pi /a$ [Figs.~\ref{fig:Supercell_profiles}(g)-(l)]. The numerical results introduced are extracted for the plane at $z = 6$~mm corresponding to half the height of the bianisotropic resonators. The time-averaged Poynting vector is calculated as $\mathbf{S} = 0.5[\mathbf{E} \times \mathbf{H}^*]$, where $\mathbf{E}$ and $\mathbf{H}$ denote complex three-component vectors of electric and magnetic fields, respectively.

It is seen that each of the magnetic field profiles in Fig.~\ref{fig:Supercell_profiles} features the concentration of the fields in the resonators neighboring the interface. Non-zero magnetic-field amplitudes in the resonators at the outer boundaries of the supercell correspond to the additional interface mentioned above, which is present due to the periodic boundary conditions. Bulk resonators feature slightly increased magnetic fields $H_z$ at the frequency $f = 2.51997$~GHz [Fig.~\ref{fig:Supercell_profiles}(c),(e)] compared to the rest of the frequencies considered, since the frequency of the associated interface state is close to the upper bulk band.

\begin{figure*}[tbp]
   \includegraphics[width=18cm]{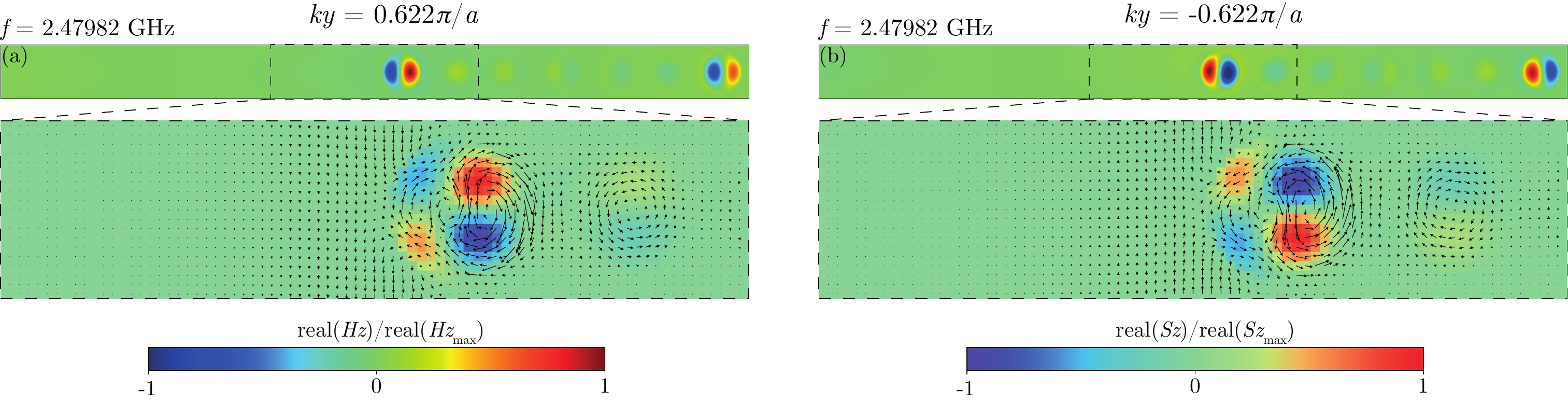}
   \caption{Numerically obtained distributions of the $H_z$ magnetic field component and zoomed-in region with normal (shown by color) and tangential (shown by black arrows) components of the Poynting vector for the supercell consisting of a single domain and free space for wave numbers (a) $k_y = 0.622 \pi /a$ and (b) $k_y = -0.622 \pi /a$ at frequency $f=2.47982$~GHz.}
   \label{fig:Supercell_vacuum_profiles}
\end{figure*}

Finally, the Poynting vector plots demonstrate the absence of directed energy flows along the interface (i.e., the localized interface states) at frequency $f=2.44401$~GHz [Figs.~\ref{fig:Supercell_profiles}(a),(d)]. At the same time, for the states at frequency $f = 2.50514$~GHz, the energy flows are observed in opposite directions along the interface that correspond to the pseudospin-up ($k_y = 0.889\pi/a$) and pseudospin-down ($k_y = -0.889\pi/a$) modes [Figs.~\ref{fig:Supercell_profiles}(b),(e)]. The same holds for the states at frequency $f=2.51997$~GHz [Figs.~\ref{fig:Supercell_profiles}(c),(f)] corresponding, in turn, to the pseudospin-down ($k_y = 0.889\pi/a$) and pseudospin-up ($k_y = -0.889\pi/a$) modes featuring less pronounced energy flows along the interface compared to Figs.~\ref{fig:Supercell_profiles}(b),(e). These results support the existence of pseudospin-momentum locked edge states featuring unidirectional propagation along the interface between two square lattices with different signs of bianisotropy.

Magnetic field profiles and Poynting vector distributions have also been extracted from numerical simulations for the supercell consisting of a single domain and the vacuum medium ($d \to \infty$) for wave numbers $k_y=\pm 0.622 \pi /a$ at the frequency $f=2.47982$~GHz associated with the edge state. Similarly to the supercell with two opposite domains, the $H_z$ component concentrates in the resonators near the interfaces, Fig.~\ref{fig:Supercell_vacuum_profiles}. At the same time, the tangential components of the Poynting vector demonstrate opposite chiralities for the positive and negative wave number values, as shown in the insets in Fig.~\ref{fig:Supercell_vacuum_profiles}.

\section*{Supplementary Note 3. Robustness to structural disorder}

\begin{figure*}[tbp]
    \includegraphics[width=12cm]{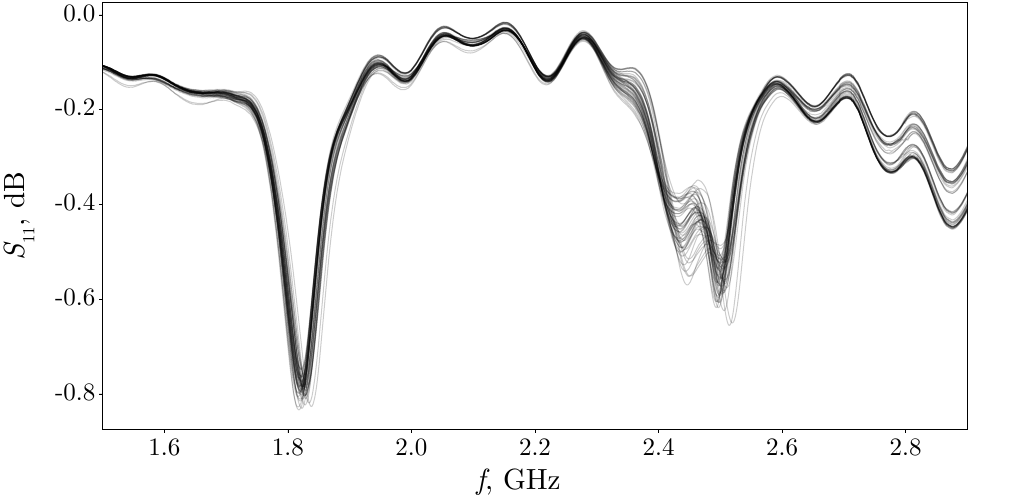}
    \caption{Experimentally measured $S_{11}$ parameters spectra for $50$ bianisotropic resonators.}
    \label{fig:Experimental_S11}
\end{figure*}

\textit{The resonant frequency.} According to the numerical results, the considered dielectric bianisotropic resonators possess two resonant peaks at frequencies $f = 2.39$~GHz and $f = 2.53$~GHz corresponding to the coupled electric and magnetic dipole moments. However, the actual resonant frequencies of the resonators in the experimental setup deviate from the idealized values due to manufacturing imperfections. Thus, we measure $S_{11}$ parameters with the help of the loop antenna having diameter $12$~mm for the $50$ randomly chosen resonators. The position of the loop antenna is the same as in the numerical simulations (inset in Fig.~2(a) in the main text). These measurements are carried out using a PLANAR~C4490 vector network analyzer. The obtained $S_{11}$ spectra are shown in Fig.~\ref{fig:Experimental_S11}. To evaluate the resonances distribution, we focus on the second resonant peak considering the range from $2.475$~GHz to $2.600$~GHz. The average frequency of the resonant peak in the mentioned frequency range is $2.501$~GHz, and the frequency deviation for different resonators ranges from $2.493$ to $2.520$~GHz. Within the experiments, we arrange the resonators in the following way: the greater the distance from a considered resonator and the interface for two-domain configurations or the studied boundary in the single-domain cases, the larger the frequency detuning of the resonator.

\begin{figure*}[tbp]
    \includegraphics[width=17cm]{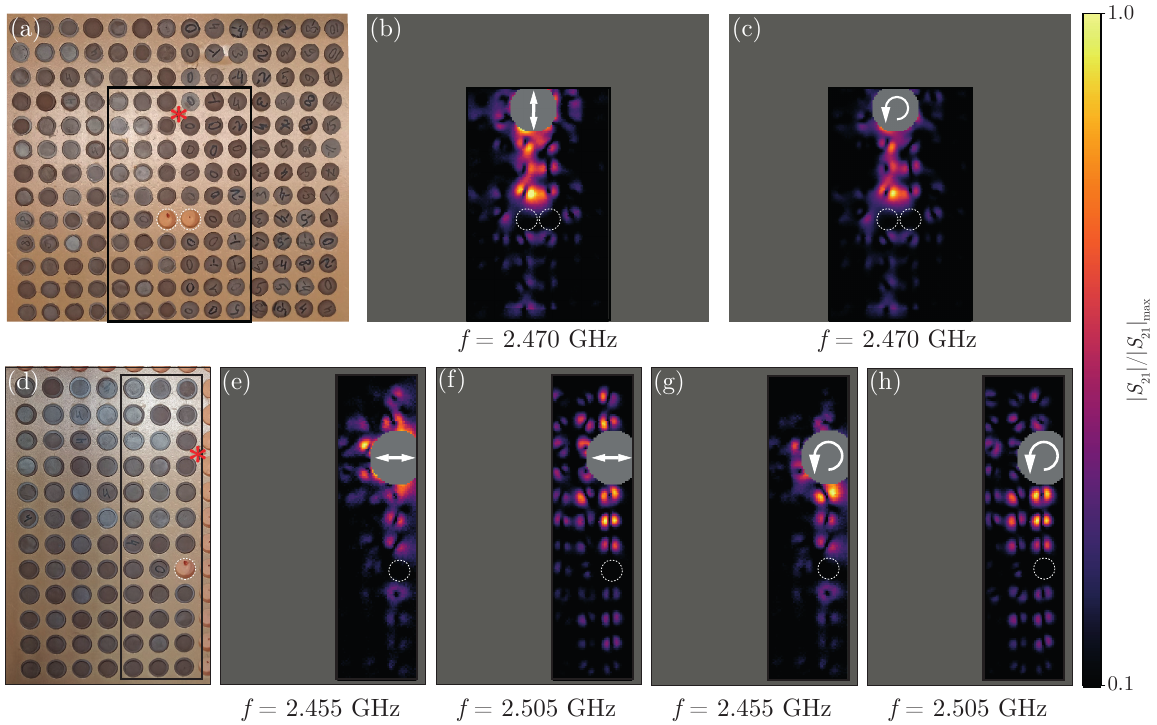}
    \caption{(a) Photo of the experimental setup featuring a linear domain wall with an extracted pair of resonators at the interface. The black frame shows the scanning area and the red asterisk denotes the position of a subwavelength source dipole antenna. (b),(c) The measured transmission coefficient profiles at the frequency $f = 2.470$~GHz for linear and circular source polarizations, respectively. (d) Photo of the one-domain configuration without a resonator at the boundary. The black frame corresponds to the scanning area and the red asterisk denotes the position of a source dipole subwavelength antenna. (e),(f) Experimentally measured magnetic field profiles in case of linear polarization excitation at the frequency of the edge state $f = 2.455$~GHz and scattering in the bulk $f = 2.505$~GHz, respectively. (g),(h) The same as panels (e),(f), but for the left-handed polarization excitation. The white dashed circles shows the position of the removed resonator.}
    \label{fig:Disorder}
\end{figure*}

\textit{Interface between two domains.} To introduce imperfections in the two-domain system with a linear interface, we consider the structure of $14 \times 13$ resonators and remove a pair of resonators, one at each side of the interface, Fig.~\ref{fig:Disorder}(a). The measurement results for linearly and circularly polarized sources are shown in Figs.~\ref{fig:Disorder}(b)-(c). In experiments, we measure near fields only in the area of interest to optimize the scanning time. Experimental results demonstrate that the edge state continues to localize at the interface. However, the magnetic field amplitude drastically decreases after the imperfection, demonstrating a strong character of such a defect.

\textit{Interface between the domain and free space.} Similarly to the interface configuration with disorder discussed before, we remove one resonator at the edge of the single-domain configuration, Fig.~\ref{fig:Disorder}(d). The magnetic field profiles at the frequencies of the edge states $f = 2.455$~GHz [Figs.~\ref{fig:Disorder}(e),(g)] and upper bulk region $f = 2.505$~GHz [Figs.~\ref{fig:Disorder}(f),(h)] demonstrate that edge state cannot propagate through such a defect, in contrast to bulk modes.

\section*{Supplementary Note 4. Bloch Hamiltonian derivation}

To derive the Bloch Hamiltonian describing a square lattice of bianisotropic resonators, we apply the Green's function approach from Ref.~\cite{2019_Gorlach}. First, we establish the following assumptions concerning the frequencies of electromagnetic hybridization ($2.39$ - $2.53$~GHz) obtained from numerical simulations (Fig.~1 in the main text):
\begin{enumerate}
    \item The considered system may be represented as a two-dimensional array of point magnetic and electric dipoles. For simplicity, we assume the resulting electric- and magnetic-field amplitudes in the resonators to be equal.

    \item In the frequency range of the bianisotropic modes, the resonators possess electric and magnetic dipoles in the $xy$ plane only, as the height of the resonators is considerably smaller than their diameter, resulting in a lower frequency of the magnetic dipole mode and a larger frequency of the electric dipole mode along the $z$ axis compared to the respective modes in the $(xy)$ plane.

    \item Each resonator interacts only with its nearest and next-nearest neighbors, i.e., with the resonators within the second coordination sphere.
\end{enumerate}

The components of the electric ($p$) and magnetic ($m$) dipole moments are related to the components of the electric and magnetic fields of the resonator at the lattice site $(n,l)$ via the polarizability tensor $\hat{\alpha}$:
\begin{equation}
    \begin{pmatrix}
    p_x^{nl}\\
    p_y^{nl}\\
    m_x^{nl}\\
    m_y^{nl}
    \end{pmatrix}
    = 
    \widehat{\alpha}
    \begin{pmatrix}
    E_x^{nl}\\
    E_y^{nl}\\
    H_x^{nl}\\
    H_y^{nl}
    \end{pmatrix}.
    \label{Dipole_moments}
\end{equation}
The polarizability tensor can be decomposed into electric $\hat{\alpha}^{ee}$, magnetic $\hat{\alpha}^{mm}$, electro-magnetic $\hat{\alpha}^{em}$, and magneto-electric $\hat{\alpha}^{me}$ parts, each being a $[2 \times 2]$ matrix:
\begin{equation}
    \widehat{\alpha} =
    \begin{pmatrix}
    \widehat{\alpha}^{ee} & \widehat{\alpha}^{em}\\
    \widehat{\alpha}^{me} & \widehat{\alpha}^{mm}
    \end{pmatrix}.
\end{equation}

We assume that the polarizabilities of the magnetic and electric dipoles are equal, while the polarizability along $z$ axis is neglected:
\begin{equation}
    \widehat{\alpha}^{ee} = \widehat{\alpha}^{mm} =
    \begin{pmatrix}
        \beta & 0 \\
        0 & \beta \\
    \end{pmatrix}.
\end{equation}
The bianisotropic parts of the polarizability tensor are defined as
\begin{equation}
    \widehat{\alpha}^{em} = \widehat{\alpha}^{me} =
    \begin{pmatrix}
        0 & {\rm i} \chi\\
        -{\rm i} \chi & 0\\
    \end{pmatrix}, 
\end{equation}
where $\chi$ is the coupling of the electric and magnetic dipoles. Thus, the polarizability tensor has the following form:
\begin{equation}
    \widehat{\alpha} = 
    \begin{pmatrix}
        \beta & 0 & 0 & {\rm i} \chi\\
        0 & \beta & -{\rm i} \chi & 0\\
        0 & {\rm i} \chi & \beta & 0\\
        -{\rm i} \chi & 0 & 0 & \beta
    \end{pmatrix}.
\end{equation}

Now, the electric and magnetic field from Eq.~\eqref{Dipole_moments} can be derived with the help of inverse polarizability tensor:
\begin{equation}
    \widehat{\alpha}^{-1}
    \begin{pmatrix}
    p_x^{nl}\\
    p_y^{nl}\\
    m_x^{nl}\\
    m_y^{nl}
    \end{pmatrix}
    =
    \begin{pmatrix}
        u & 0 & 0 & -{\rm i}v\\
        0 & u & {\rm i}v & 0\\
        0 & -{\rm i}v & u & 0\\
        {\rm i}v & 0 & 0 & u
    \end{pmatrix}
        \begin{pmatrix}
    p_x^{nl}\\
    p_y^{nl}\\
    m_x^{nl}\\
    m_y^{nl}
    \end{pmatrix}
    =
    \begin{pmatrix}
    E_x^{nl}\\
    E_y^{nl}\\
    H_x^{nl}\\
    H_y^{nl}
    \end{pmatrix}
    \label{Fields_by_dipole_moments},
\end{equation}
where $u = \dfrac{\beta}{\beta^2-\chi^2}$, and $v = \dfrac{\chi}{\beta^2-\chi^2}$.

On the other hand, the electromagnetic fields can be evaluated via the Green's functions for point dipoles:
\begin{align}
    \mathbf{E}^{nl} & = \widehat{G}^{ee}(r_{ij}, k_0)\mathbf{p}^{ij} + \widehat{G}^{em}(r_{ij}, k_0)\mathbf{m}^{ij}, \\
    \mathbf{H}^{nl} & = \widehat{G}^{me}(r_{ij}, k_0)\mathbf{p}^{ij} +\widehat{G}^{mm}(r_{ij}, k_0)\mathbf{m}^{ij},
\end{align}
where $r_{ij}$ is the distance between the resonators with coordinates $(n,l)$ and $(i,j)$, $k_0$ is the wave number, and the contribution of the electric dipole in the electric field is the same as the contribution of the magnetic dipole in the magnetic field $\widehat{G}^{mm}(r_{ij}, k_0) = \widehat{G}^{ee}(r_{ij}, k_0)$, while Green's functions for the bianisotropic parts are related as $\widehat{G}^{em}(r_{ij}, k_0) = -\widehat{G}^{me}(r_{ij}, k_0)$. Dyadic Green's function components can be expressed in the following manner:
\begin{align}
    \widehat{G}^{ee}_{\zeta\eta} & = (\partial_{\zeta} \partial_{\eta} + k^2_0\delta_{\zeta\eta})\dfrac{e^{{\rm i} k_0 r_{ij}}}{r_{ij}},
    \label{eq:Gee} \\
    \widehat{G}^{em}_{\zeta\eta} & = {\rm i} k_0 \epsilon_{\zeta\eta\kappa}\partial_{\kappa}\dfrac{e^{{\rm i} k_0 r_{ij}}}{r_{ij}},
    \label{eq:Gem}
\end{align}
where $\delta_{\zeta,\eta}$ is the Kronecker delta, $\varepsilon_{\zeta,\eta, \kappa}$ is Levi-Civita symbol, and $\partial_\zeta = \dfrac{\partial}{\partial \zeta}$. The parameters $\zeta$, $\eta$, $\kappa$ take the values $\{ x,y,z\}$.

If the distance between the resonators $r_{ij} = r = \sqrt{x^2 + y^2}$, then, according to Eq.~\eqref{eq:Gee} and Eq.~\eqref{eq:Gem}, the components of the dyadic Green's functions are
\begin{align}
     \widehat{G}^{ee}_{xx} & = \dfrac{e^{{\rm i} k_0 r}}{r^3}(\dfrac{x^2}{r^2}(3-3 {\rm i} k_0 r - k_0^2 r^2)+{\rm i} k_0 r + k^2_0 r^2 - 1), \label{Gee_xx_general} \\
    \widehat{G}^{ee}_{yy} & = \dfrac{e^{{\rm i} k_0 r}}{r^3}(\dfrac{y^2}{r^2}(3-3 {\rm i} k_0 r - k_0^2 r^2)+{\rm i} k_0 r + k^2_0 r^2 - 1), \label{Gee_yy_general} \\
    \widehat{G}^{ee}_{xy} & = \widehat{G}^{ee}_{yx}= \dfrac{e^{{\rm i} k_0 r}}{r^3}(\dfrac{xy}{r^2}(3-3 {\rm i} k_0 r - k_0^2 r^2)), \label{Gee_xy_general}
\end{align}
where $x$ and $y$ are the distances along the $x$ and $y$ axes, respectively. As the in-plane distance $r$ is independent of the component $\kappa = z$, the function $\widehat{G}^{em}_{\zeta,\eta} = 0$ for any combination of $\zeta$ and $\eta$ for the considered system.

Next, we assume that the field within the resonator with coordinates $(n,l)$ is produced by the eight closest resonators $(i,j)$ in the two-dimensional resonator array:
\begin{align}
    \mathbf{E}^{nl} = &\sum_{i = n \pm 1, j = l}\widehat{G}^{ee}(r_{ij},k_0)\mathbf{p}^{ij}  + \sum_{i = n, j = l \pm 1}\widehat{G}^{ee}({r}_{ij},k_0)\mathbf{p}^{ij} + \sum_{\substack{i = n + 1, j = l + 1\\
    i = n - 1, j = l - 1}}\widehat{G}^{ee}({r}_{ij},k_0)\mathbf{p}^{ij} + \sum_{\substack{i = n - 1, j = l + 1\\
    i = n + 1, j = l - 1}} \widehat{G}^{ee}({r}_{ij},k_0)\mathbf{p}^{ij} + \nonumber \\
     + &\sum_{i = n \pm 1, j = l}\widehat{G}^{em}({r}_{ij},k_0)\mathbf{m}^{ij} + \sum_{i = n, j = l \pm 1}\widehat{G}^{em}({r}_{ij},k_0)\mathbf{m}^{ij} + \sum_{\substack{i = n + 1, j = l + 1\\
    i = n - 1, j = l - 1}}\widehat{G}^{em}({r}_{ij},k_0)\mathbf{m}^{ij} + \sum_{\substack{i = n - 1, j = l + 1\\
    i = n + 1, j = l - 1}}\widehat{G}^{em}({r}_{ij},k_0)\mathbf{m}^{ij},
     \label{eq:E_Green_total}
\end{align}

\begin{align}
    \mathbf{H}^{nl} = &\sum_{i = n \pm 1, j = l}\widehat{G}^{ee}({r}_{ij},k_0)\mathbf{m}^{ij} + \sum_{i = n, j = l \pm 1}\widehat{G}^{ee}({r}_{ij},k_0)\mathbf{m}^{ij} + \sum_{\substack{i = n + 1, j = l + 1\\
    i = n - 1, j = l - 1}}\widehat{G}^{ee}({r}_{ij},k_0)\mathbf{m}^{ij} + \sum_{\substack{i = n - 1, j = l + 1\\
    i = n + 1, j = l - 1}}\widehat{G}^{ee}({r}_{ij},k_0)\mathbf{m}^{ij}\nonumber \\
    - &\sum_{i = n \pm 1, j = l}\widehat{G}^{em}({r}_{ij},k_0)\mathbf{p}^{ij} - \sum_{i = n, j = l \pm 1}\widehat{G}^{em}({r}_{ij},k_0)\mathbf{p}^{i,j} - \sum_{\substack{i = n + 1, j = l + 1\\
    i = n - 1, j = l - 1}}\widehat{G}^{em}({r}_{ij},k_0)\mathbf{p}^{i,j} - \sum_{\substack{i = n - 1, j = l + 1\\
    i = n + 1, j = l - 1}}\widehat{G}^{em}({r}_{ij},k_0)\mathbf{p}^{i,j}.
    \label{eq:H_Green_total}
\end{align}

Then, we rewrite Eqs.~\eqref{Gee_xx_general},\eqref{Gee_yy_general}, and \eqref{Gee_xy_general} taking into account the lattice constant $a$ (by substituting $r = a$) and the coordinates $i, j, n, l$:
\begin{equation}
    \widehat{G}^{ee}_{xx} = \dfrac{e^{{\rm i} k_0 a}}{a^3}\big((i - n)^2(3-3 {\rm i} k_0 a - k_0^2 a^2)+{\rm i} k_0 a + k^2_0 a^2 - 1\big),
    \label{Gee_xx}
\end{equation}
\begin{equation}
    \widehat{G}^{ee}_{yy} = \dfrac{e^{{\rm i} k_0 a}}{a^3}\big((j - l)^2(3-3 {\rm i} k_0 a - k_0^2 a^2)+{\rm i} k_0 a + k^2_0 a^2 - 1\big),
    \label{Gee_yy}
\end{equation}
\begin{equation}
    \widehat{G}^{ee}_{xy} = \widehat{G}^{ee}_{yx}= \dfrac{e^{{\rm i} k_0 a}}{a^3}\big((i-n)(j - l)(3-3 {\rm i} k_0 a - k_0^2 a^2)\big).
    \label{Gee_xy}
\end{equation}

Now, these functions will be different for different pairs of neighboring resonators. If $i = n \pm 1$ and $j = l$, then the dyadic Green's function takes the following form:
\begin{align*}
   \widehat{G}^{ee}_{xx} & = A = 2\dfrac{e^{{\rm i} k_0 a}}{a^3}(1-{\rm i} k_0 a),\\
   \widehat{G}^{ee}_{yy} & = B = \dfrac{e^{{\rm i} k_0 a}}{a^3}({\rm i} k_0 a + k^2_0 a^2 - 1),\\
   \widehat{G}^{ee}_{xy} & = \widehat{G}^{ee}_{yx} = 0.
\end{align*}
If $i = n$ and $j = l \pm 1$:
\begin{align*}
   \widehat{G}^{ee}_{xx} & = B = \dfrac{e^{{\rm i} k_0 a}}{a^3}({\rm i} k_0 a + k^2_0 a^2 - 1),\\
   \widehat{G}^{ee}_{yy} & = A = 2\dfrac{e^{{\rm i} k_0 a}}{a^3}(1-{\rm i} k_0 a),\\
   \widehat{G}^{ee}_{xy} & = \widehat{G}^{ee}_{yx} = 0. 
\end{align*}

If $i = n + 1$ and $j = l + 1$ or $i = n - 1$ and $j = l - 1$ (the upper right and the lower left diagonal neighbors), the distance between the considered nodes is $r = \sqrt{2}a$, and the dyadic Green's functions read
\begin{align*}
   \widehat{G}^{ee}_{xx} & = \widehat{G}^{ee}_{yy} = C = \dfrac{e^{{\rm i} k_0 \sqrt{2} a}}{4 \sqrt{2} a^3}(1 - {\rm i} k_0 \sqrt{2} a + 2k^2_0 a^2),\\
   \widehat{G}^{ee}_{xy} & = \widehat{G}^{ee}_{yx} = D = 
   \dfrac{e^{{\rm i} k_0 \sqrt{2} a}}{4 \sqrt{2} a^3}(3 - {\rm i} 3 k_0 \sqrt{2} a - 2k^2_0 a^2).
\end{align*}

If $i = n - 1$ and $j = l + 1$ or $i = n + 1$ and $j = l - 1$ (the upper left and the lower right diagonal neighbors) the dyadic Green functions $\widehat{G}^{ee}_{xx} = \widehat{G}^{ee}_{yy} = C$, and $\widehat{G}^{ee}_{xy} = \widehat{G}^{ee}_{yx} = -D$.

Finally, we can relate the electric and magnetic moments at the point with coordinates $(n,l)$ defined by Eq.~\eqref{Fields_by_dipole_moments} and at the point $(i,j)$ given by Eqs.~\eqref{eq:E_Green_total},\eqref{eq:H_Green_total} using the obtained dyadic Green's functions $A$, $B$, $C$, and $D$ for the certain pairs of the neighbors:
\begin{align}
    \begin{split}
    \widehat{\alpha}^{-1}
    \begin{pmatrix}
    p_{x}^{nl}\\
    p_{y}^{nl}\\
    m_{x}^{nl}\\
    m_{y}^{nl}
    \end{pmatrix}
    =
    & \sum_{i = n \pm 1, j = l}
    \begin{pmatrix}
    A & 0 & 0 & 0 \\
    0 & B & 0 & 0 \\
    0 & 0 & A & 0 \\
    0 & 0 & 0 & B \\
    \end{pmatrix}
     \begin{pmatrix}
    p_{x}^{ij}\\
    p_{y}^{ij}\\
    m_{x}^{ij}\\
    m_{y}^{ij}
    \end{pmatrix}
    +
    \sum_{i = n, j = l \pm 1}
    \begin{pmatrix}
    B & 0 & 0 & 0 \\
    0 & A & 0 & 0 \\
    0 & 0 & B & 0 \\
    0 & 0 & 0 & A \\
    \end{pmatrix}
     \begin{pmatrix}
    p_{x}^{ij}\\
    p_{y}^{ij}\\
    m_{x}^{ij}\\
    m_{y}^{ij} + \\
    +
    \end{pmatrix}
    + \\
    +
    & \sum_{\substack{i = n + 1, j = l + 1\\
    i = n - 1, j = l - 1}}
    \begin{pmatrix}
    C & D & 0 & 0 \\
    D & C & 0 & 0 \\
    0 & 0 & C & D \\
    0 & 0 & D & C \\
    \end{pmatrix}
     \begin{pmatrix}
    p_{x}^{ij}\\
    p_{y}^{ij}\\
    m_{x}^{ij}\\
    m_{y}^{ij}
    \end{pmatrix}+
    \sum_{\substack{i = n - 1, j = l + 1\\
    i = n + 1, j = l - 1}}
    \begin{pmatrix}
    C & -D & 0 & 0 \\
    -D & C & 0 & 0 \\
    0 & 0 & C & -D \\
    0 & 0 & -D & C \\
    \end{pmatrix}
     \begin{pmatrix}
    p_{x}^{ij}\\
    p_{y}^{ij}\\
    m_{x}^{ij}\\
    m_{y}^{ij}
    \end{pmatrix}.
    \end{split}
    \label{eq:Main}
\end{align} 

According to Bloch theorem, the electric and magnetic dipole moments of a certain resonator are related with the ones for its nearest neighbors as
\begin{equation*}
     \begin{pmatrix}
    p_{x}^{i+1,j}\\
    p_{y}^{i+1,j}\\
    m_{x}^{i+1,j}\\
    m_{y}^{i+1,j}
    \end{pmatrix}
    = e^{{\rm i} k_x}
         \begin{pmatrix}
    p_{x}^{ij}\\
    p_{y}^{ij}\\
    m_{x}^{ij}\\
    m_{y}^{ij}
    \end{pmatrix}
    ,
     \begin{pmatrix}
    p_{x}^{i-1,j}\\
    p_{y}^{i-1,j}\\
    m_{x}^{i-1,j}\\
    m_{y}^{i-1,j}
    \end{pmatrix}
    = e^{-{\rm i} k_x}
         \begin{pmatrix}
    p_{x}^{ij}\\
    p_{y}^{ij}\\
    m_{x}^{ij}\\
    m_{y}^{ij}
    \end{pmatrix},
\end{equation*}

\begin{equation*}
     \begin{pmatrix}
    p_{x}^{i,j+1}\\
    p_{y}^{i,j+1}\\
    m_{x}^{i,j+1}\\
    m_{y}^{i,j+1}
    \end{pmatrix}
    = e^{{\rm i} k_y}
         \begin{pmatrix}
    p_{x}^{ij}\\
    p_{y}^{ij}\\
    m_{x}^{ij}\\
    m_{y}^{ij}
    \end{pmatrix}
    ,
     \begin{pmatrix}
    p_{x}^{i,j-1}\\
    p_{y}^{i,j-1}\\
    m_{x}^{i,j-1}\\
    m_{y}^{i,j-1}
    \end{pmatrix}
    = e^{-{\rm i} k_y}
    \begin{pmatrix}
    p_{x}^{ij}\\
    p_{y}^{ij}\\
    m_{x}^{ij}\\
    m_{y}^{ij}
    \end{pmatrix},
\end{equation*}

\begin{equation*}
     \begin{pmatrix}
    p_{x}^{i+1,j+1}\\
    p_{y}^{i+1,j+1}\\
    m_{x}^{i+1,j+1}\\
    m_{y}^{i+1,j+1}
    \end{pmatrix}
    = e^{{\rm i} (k_x + k_y)}
         \begin{pmatrix}
    p_{x}^{ij}\\
    p_{y}^{ij}\\
    m_{x}^{ij}\\
    m_{y}^{ij}
    \end{pmatrix}
    ,
     \begin{pmatrix}
    p_{x}^{i-1,j-1}\\
    p_{y}^{i-1,j-1}\\
    m_{x}^{i-1,j-1}\\
    m_{y}^{i-1,j-1}
    \end{pmatrix}
    = e^{-{\rm i} (k_x + k_y)}
    \begin{pmatrix}
    p_{x}^{ij}\\
    p_{y}^{ij}\\
    m_{x}^{ij}\\
    m_{y}^{ij}
    \end{pmatrix},
\end{equation*}

\begin{equation*}
     \begin{pmatrix}
    p_{x}^{i+1,j-1}\\
    p_{y}^{i+1,j-1}\\
    m_{x}^{i+1,j-1}\\
    m_{y}^{i+1,j-1}
    \end{pmatrix}
    = e^{{\rm i} (k_x - k_y)}
         \begin{pmatrix}
    p_{x}^{ij}\\
    p_{y}^{ij}\\
    m_{x}^{ij}\\
    m_{y}^{ij}
    \end{pmatrix}
    ,
     \begin{pmatrix}
    p_{x}^{i-1,j+1}\\
    p_{y}^{i-1,j+1}\\
    m_{x}^{i-1,j+1}\\
    m_{y}^{i-1,j+1}
    \end{pmatrix}
    = e^{-{\rm i} (k_x - k_y)}
    \begin{pmatrix}
    p_{x}^{ij}\\
    p_{y}^{ij}\\
    m_{x}^{ij}\\
    m_{y}^{ij}
    \end{pmatrix}.
\end{equation*}

The left hand side of Eq.~\eqref{eq:Main} can be rewritten in the form
\begin{equation}
    \begin{pmatrix}
    u & 0 & 0 & -{\rm i}v\\
    0 & u & {\rm i}v & 0\\
    0 & -{\rm i}v & u & 0\\
    {\rm i}v & 0 & 0 & u
    \end{pmatrix}
    \begin{pmatrix}
    p_{x}^{nl}\\
    p_{y}^{nl}\\
    m_{x}^{nl}\\
    m_{y}^{nl}
    \end{pmatrix}
    = u
        \begin{pmatrix}
        1 & 0 & 0 & 0\\
        0 & 1 & 0 & 0\\
        0 & 0 & 1 & 0\\
        0 & 0 & 0 & 1
    \end{pmatrix}
    \begin{pmatrix}
    p_{x}^{nl}\\
    p_{y}^{nl}\\
    m_{x}^{nl}\\
    m_{y}^{nl}
    \end{pmatrix}
    + {\rm i} v
    \begin{pmatrix}
    0 & 0 & 0 & -1\\
    0 & 0 & 1 & 0\\
    0 & -1 & 0 & 0\\
    1 & 0 & 0 & 0
    \end{pmatrix}
    \begin{pmatrix}
    p_{x}^{nl}\\
    p_{y}^{nl}\\
    m_{x}^{nl}\\
    m_{y}^{nl}
    \end{pmatrix}.
\end{equation}

In the following, we rely on the approximations $u = \dfrac{\omega - \omega_{0}}{\rm const}$, where $\omega_{0}$ is the resonant frequency of a resonator, and $v = \rm const$~\cite{2019_Gorlach}. To convert the equation to dimensionless values, we set $q = ak_0$, $\mu = 2 v a^3$, and $E = 2a^3 \dfrac{\omega - \omega_{0}}{\rm const}$. Then,
\begin{equation}
    u \hat{I}_4
    \begin{pmatrix}
    p_{x}^{ij}\\
    p_{y}^{ij}\\
    m_{x}^{ij}\\
    m_{y}^{ij}
    \end{pmatrix}
    + {\rm i} v
    \begin{pmatrix}
    0 & 0 & 0 & -1\\
    0 & 0 & 1 & 0\\
    0 & -1 & 0 & 0\\
    1 & 0 & 0 & 0
    \end{pmatrix}
    \begin{pmatrix}
    p_{x}^{ij}\\
    p_{y}^{ij}\\
    m_{x}^{ij}\\
    m_{y}^{ij}
    \end{pmatrix}
    \equiv \dfrac{E}{2 a^3} \hat{I}_4
    \begin{pmatrix}
    p_{x}^{ij}\\
    p_{y}^{ij}\\
    m_{x}^{ij}\\
    m_{y}^{ij}
    \end{pmatrix}
    + \dfrac{{\rm i} \mu}{2 a^3}
    \begin{pmatrix}
    0 & 0 & 0 & -1\\
    0 & 0 & 1 & 0\\
    0 & -1 & 0 & 0\\
    1 & 0 & 0 & 0
    \end{pmatrix}
    \begin{pmatrix}
    p_{x}^{ij}\\
    p_{y}^{ij}\\
    m_{x}^{ij}\\
    m_{y}^{ij}
    \end{pmatrix},
\end{equation}
where $\hat{I}_{4}$ is the $[4 \times 4]$ unity matrix. Finally, Eq.~\eqref{eq:Main} can be represented as:
\begin{align}
    \Biggl[
    & 2a^3(e^{{\rm i} k_x} + e^{-{\rm i} k_x})
    \begin{pmatrix}
    A & 0 & 0 & 0 \\
    0 & B & 0 & 0 \\
    0 & 0 & A & 0 \\
    0 & 0 & 0 & B \\
    \end{pmatrix}
    +2a^3(e^{{\rm i} k_y} + e^{-{\rm i} k_y})
    \begin{pmatrix}
    B & 0 & 0 & 0 \\
    0 & A & 0 & 0 \\
    0 & 0 & B & 0 \\
    0 & 0 & 0 & A \\
    \end{pmatrix}
    +2 a^3 (e^{{\rm i}(k_x + k_y)}+e^{-{\rm i}(k_x + k_y)})
    \begin{pmatrix}
    C & D & 0 & 0 \\
    D & C & 0 & 0 \\
    0 & 0 & C & D \\
    0 & 0 & D & C \\
    \end{pmatrix} + \nonumber \\
    + & 2 a^3 (e^{{\rm i}(k_x - k_y)}+e^{-{\rm i}(k_x - k_y)})
    \begin{pmatrix}
    C & -D & 0 & 0 \\
    -D & C & 0 & 0 \\
    0 & 0 & C & -D \\
    0 & 0 & -D & C \\
    \end{pmatrix} 
    - {\rm i}\mu
    \begin{pmatrix}
    0 & 0 & 0 & -1\\
    0 & 0 & 1 & 0\\
    0 & -1 & 0 & 0\\
    1 & 0 & 0 & 0
    \end{pmatrix}
    \Biggr]
    \begin{pmatrix}
    p_{x}^{ij}\\
    p_{y}^{ij}\\
    m_{x}^{ij}\\
    m_{y}^{ij}
    \end{pmatrix} 
    = E \hat{I}_4
    \begin{pmatrix}
    p_{x}^{ij}\\
    p_{y}^{ij}\\
    m_{x}^{ij}\\
    m_{y}^{ij}
    \end{pmatrix}.
    \label{eq:pre_Hamiltonian}
\end{align}

In the following, we consider the near-field contribution in the Green function exclusively. Also, we assume that $q \ll 1$, because $\lambda \gg a$, where $\lambda$ is wavelength corresponding to the frequencies around $2.5$~GHz. Thus, the exponent approximately equals $\exp^{{\rm i} q} \approx 1$. Finally, $a^{3}A \approx 2$, $a^{3}B \approx -1$, $a^{3}C \approx 1/(4 \sqrt{2})$, $a^{3}D \approx 3/(4 \sqrt{2})$, $e^{{\rm i} k} + e^{-{\rm i} k} = 2 \cos k$, and the derived Bloch Hamiltonian takes the form
\begin{align}
    & \widehat{H}(k_x,k_y) =
    \begin{pmatrix}
    h_{11} & h_{12} & 0 & {\rm i} \mu\\
    h_{21} & h_{22} & -{\rm i} \mu & 0\\
    0 & {\rm i} \mu & h_{11} & h_{12}\\
    -{\rm i} \mu & 0 & h_{21} & h_{22} 
    \end{pmatrix}, \\
    & h_{11} = 8 \cos k_x - 4 \cos k_y + \dfrac{1}{\sqrt{2}}(\cos (k_x + k_y) + \cos (k_x - k_y)), \\
    & h_{12} = h_{21} = \dfrac{3}{\sqrt{2}}(\cos (k_x + k_y) - \cos (k_x - k_y)),\\
    & h_{22} = 8 \cos k_y - 4 \cos k_x + \dfrac{1}{\sqrt{2}}(\cos (k_x + k_y) + \cos (k_x - k_y)).
\end{align}

For further topological invariant evaluation, we rewrite the derived Hamiltonian in the pseudospin basis $\ket{\psi}^{\prime} = (p_{x}+m_{x}, p_{y}+m_{y}, p_{x}-m_{x}, p_{y}-m_{y})^{\rm T}$~\cite{2017_Slobozhanyuk}. Thus, the unitary transform matrix $\hat{U}$ is the following:
\begin{equation}
    \hat{U} = \dfrac{1}{\sqrt{2}}
    \begin{pmatrix}
        1 & 0 & 1 & 0\\
        0 & 1 & 0 & 1\\
        1 & 0 & -1 & 0\\
        0 & 1 & 0 & -1
    \end{pmatrix}.
\end{equation}

Indeed, $\ket{\psi}^{\prime} = \hat{U}\ket{\psi}$, where $\ket{\psi}$ is in the old basis $\ket{\psi} = (p_x, p_y, m_x, m_y)^{\rm T}$, and $\hat{H}^{\prime}(k_{x},k_{y}) = \hat{U}\hat{H}(k_{x},k_{y})\hat{U}^{\dag}$ takes the following block-diagonal form:
\begin{align}
   \hat{H}(k_{x},k_{y}) & = \big( \sqrt{2} \cos k_x \cos k_y + 2 \cos k_x + 2 \cos k_y \big) \hat{I}\otimes\hat{I} + 6(\cos k_x - \cos k_y) \hat{I}\otimes\hat{\sigma_z} - 3 \sqrt{2} \sin k_x \sin k_y \hat{I}\otimes\hat{\sigma_x}- \nonumber\\ 
   & -\mu \hat{\sigma_z}\otimes\hat{\sigma_y} =  
   \begin{pmatrix}
       \hat{H}^{\uparrow} & 0\\
       0 & \hat{H}^{\downarrow}
   \end{pmatrix},
   \label{eq:spin_H} 
\end{align}
where $\hat{\sigma_{x}} = (0, 1; 1, 0)$, $\hat{\sigma_{y}} = (0, -{\rm i}; {\rm i}, 0)$, and $\hat{\sigma_{z}}=(1, 0; 0, -1)$ are Pauli matrices and $\hat{I}$ is unity matrix.

\begin{figure*}[tbp]
    \includegraphics[width=8cm]{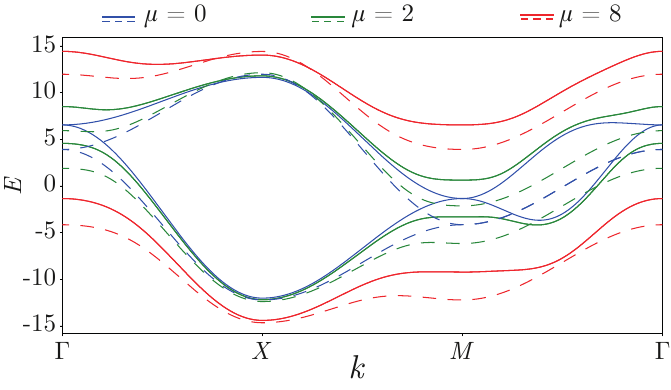}
    \caption {The dispersion diagrams for the eigenvalues $E(k)$ Eq.~\eqref{eq:Energies_NNN} within the next-nearest neighbors approximation (the solid lines) and $E^{\prime}(k)$~Eq.~\eqref{eq:E_NN} within the nearest neighbors approximation (the dashed lines) for the wave number $k=(k_{x},k_{y})$ varied along the $\Gamma-X-M-\Gamma$ trajectory in the reciprocal space for different values of the bianisotropy $\mu=0$ (the blue lines), ${\mu=2}$ (the green lines), and $\mu=8$ (the red lines), respectively.}
    \label{fig:Hamiltonian_dispersion_NN_and_NNN}
\end{figure*}

The pseudospin-up $\hat{H}^{\uparrow}$ and pseudospin-down $\hat{H}^{\downarrow}$ parts of the Hamiltonian~\eqref{eq:spin_H} feature the following bands of the eigenenergies:
\begin{align}
    E^{\uparrow(\downarrow)}_{\pm} & = 2(\cos k_x + \cos k_y) + \sqrt{2} \cos k_x \cos k_y \pm \nonumber \\
    & \pm \dfrac{3}{\sqrt{2}}\sqrt{9 + \dfrac{2}{9}\mu^2 - 16 \cos k_x \cos k_y + 3 \cos(2k_y) + 3 \cos(2k_x) + \cos(2k_x)\cos(2k_y)}.
    \label{eq:Energies_NNN}
\end{align}
These energy bands are shown by solid lines in Fig.~\ref{fig:Hamiltonian_dispersion_NN_and_NNN} for different values of the parameter $\mu$.

Eigenfunctions of the Hamiltonians $\hat{H}^{\uparrow}$ and $\hat{H}^{\downarrow}$ have the following form:
\begin{align}
     \ket{\psi}^{\uparrow}_{1} =& \big( \frac{12(\cos k_x - \cos k_y) + \sqrt{2} \sqrt{81 + 2\mu^2 - 144 \cos k_x \cos k_y + 27 \cos 2k_x + 27 \cos 2k_y + 9\cos 2k_x \cos 2k_y}}{2 {\rm i} \mu + 6 \sqrt{2} \sin k_x \sin k_y},1\big)^{\rm T}, \nonumber \\
     \ket{\psi}^{\uparrow}_{2} =& \big( \frac{12(\cos k_x - \cos k_y) - \sqrt{2} \sqrt{81 + 2\mu^2 - 144 \cos k_x \cos k_y + 27 \cos 2k_x + 27 \cos 2k_y + 9\cos 2k_x \cos 2k_y}}{2 {\rm i} \mu + 6 \sqrt{2} \sin k_x \sin k_y},1\big)^{\rm T}, \nonumber \\
      \ket{\psi}^{\downarrow}_{1} =& \big( \frac{12(\cos k_x - \cos k_y) + \sqrt{2} \sqrt{81 + 2\mu^2 - 144 \cos k_x \cos k_y + 27 \cos 2k_x + 27 \cos 2k_y + 9\cos 2k_x \cos 2k_y}}{-2 {\rm i} \mu + 6 \sqrt{2} \sin k_x \sin k_y},1\big)^{\rm T}, \nonumber \\
        \ket{\psi}^{\downarrow}_{2} =& \big( \frac{12(\cos k_x - \cos k_y) - \sqrt{2} \sqrt{81 + 2\mu^2 - 144 \cos k_x \cos k_y + 27 \cos 2k_x + 27 \cos 2k_y + 9\cos 2k_x \cos 2k_y}}{-2 {\rm i} \mu + 6 \sqrt{2} \sin k_x \sin k_y},1\big)^{\rm T}. 
\end{align}

In the nearest-neighbor approximation, when the Green functions $D=C=0$ in Eq.~\eqref{eq:pre_Hamiltonian}, the Hamiltonian $\hat{H}^{\prime}$ in pseudospin basis attains the following form:
\begin{multline*}
    \hat{H}^{\prime}(k_x, k_y) = 
    \begin{pmatrix}
        8 \cos{k_x} - 4 \cos{k_y} & {\rm i} \mu & 0 & 0\\
        -{\rm i} \mu & -4 \cos{k_x} + 8 \cos{k_y}  & 0 & 0\\
        0 & 0 & 8 \cos{k_x} - 4 \cos{k_y} & -{\rm i} \mu \\
        0 & 0 & {\rm i} \mu & -4 \cos{k_x} + 8 \cos{k_y} 
    \end{pmatrix}
   = 
    \begin{pmatrix}
        \hat{H}^{\prime \uparrow} & 0\\
        0 & \hat{H}^{\prime \downarrow}
    \end{pmatrix}, 
\end{multline*}
with eigenenergies $E^{\prime}(k_x, k_y)$ given by the relation
\begin{align}
    E^{\prime \uparrow (\downarrow)}(k_x, k_y)  = 2 (\cos{k_x} + \cos{k_y}) \pm \sqrt{36(\cos{k_x} + \cos{k_y})^2 + \mu^2} 
    \label{eq:E_NN}
\end{align}
and eigenfunctions $\ket{\psi}^{\prime \uparrow (\downarrow)}$ being the following:
\begin{align}
     \ket{\psi}^{\prime \uparrow}_{1} =& \big(-\frac{{\rm i}(-6(\cos{k_x} - \cos{k_y}) + \sqrt{36(\cos{k_x} - \cos{k_y})^2+\mu^2})}{\mu}, 1\big)^{\rm T}, \nonumber \\
     \ket{\psi}^{\prime \uparrow}_{2} =& \big(\frac{{\rm i}(6(\cos{k_x} - \cos{k_y}) + \sqrt{36(\cos{k_x} - \cos{k_y})^2+\mu^2})}{\mu}, 1\big)^{\rm T}, \nonumber \\
      \ket{\psi}^{\prime \downarrow}_{1} =& \big(\frac{{\rm i}(-6(\cos{k_x} - \cos{k_y}) + \sqrt{36(\cos{k_x} - \cos{k_y})^2+\mu^2})}{\mu}, 1\big)^{\rm T}, \nonumber \\
      \ket{\psi}^{\prime \downarrow}_{2} =& \big(-\frac{{\rm i}(6(\cos{k_x} - \cos{k_y}) + \sqrt{36(\cos{k_x} - \cos{k_y})^2+\mu^2})}{\mu}, 1\big)^{\rm T}.
      \label{eq:Eigenfunctions_NN}
\end{align}
The dispersion curves for the eigenenergies $E^{\prime}$ are plotted in Fig.~\ref{fig:Hamiltonian_dispersion_NN_and_NNN} with the dashed lines for different values of the parameter $\mu$.

It is seen that for $\mu > 0$ a bandgap opens, similar to the previously considered case with the next-nearest neighbors taken into account. However, there are two important differences. First, it is seen that a nodal line between the points $\Gamma$ and $M$ appears in the nearest-neighbor approximation for $\mu=0$ which is absent in the next-nearest neighbor approximation, and the energy bands become fourfold-degenerate in a certain direction within the Brillouin zone. Second, the Berry curvature vanishes in the nearest-neighbor approximation for arbitrary values of $\mu$, as can be demonstrated by a direct evaluation for the eigenfunctions Eq.~\eqref{eq:Eigenfunctions_NN}. Thus, such an approximation does not correctly describe topological properties of the considered system, facilitating the necessity to consider at least the diagonally-opposite next-nearest resonators.

\section*{Supplementary Note 5. Experimental evaluation of the dispersion diagram}

To characterize the dispersion of the edge states, we place the source antenna at the edge of the domain wall and measure the excited magnetic field along the interface. Then, we evaluate the dynamic structure factor (DSF) $S(k,f)$ by applying Fourier transform to the extracted array of frequency-dependent $S_{21}$-parameters measured at different spatial points:
\begin{equation}
   S(k,f) = \sum_{j=0}^{N-1} S_{21}(y_{j},f){\rm e}^{-{\rm i} ky_{j}} = \sum_{j=0}^{N-1} S_{21}(j,f){\rm e}^{-{\rm i}k\Delta j/a},
\end{equation}
where $S_{21}(y_{j},f) \equiv S_{21}(j,f)$ is the transmission coefficient at the frequency $f$ for the loop probe located at the point ${y_{j}=j\Delta}$~mm along the interface and the exciting dipole antenna located at the point $y=0$~mm, $\Delta=2.5$~mm is the scanning step, $N=185$ is the total number of scanning points, $a=37.1$~mm is the structure period, and $k \in [0, \pi]$ is the dimensionless wave number.